\documentclass{aa}  

\usepackage{hyperref}
\usepackage{graphicx}
\usepackage{txfonts}
\usepackage{amsmath}
\usepackage{gensymb}
\usepackage{xcolor}
\usepackage{subfig}
\usepackage{caption}
\usepackage{array}
\usepackage{siunitx}

\newcommand{\pkg}[1]{\texttt{#1}}

\makeatletter
\renewcommand{\fnum@table}{\textbf{Table~\thetable}}
\makeatother

\makeatletter
\renewcommand{\fnum@figure}{\textbf{Fig.~\thefigure}}
\makeatother
\usepackage{xcolor}

\usepackage{textcomp}
\begin{document}

   \title{Intrinsic alignments in the \pkg{FLAMINGO} simulations with two-point statistics}

  \author{A. Herle \inst{1}
  \and N. E. Chisari \inst{2, 1}
  \and H. Hoekstra \inst{1} 
  \and D. Navarro-Giron\'es \inst{1} 
  \and M. Schaller \inst{3, 1} 
  \and J. Schaye \inst{1}
  }

   \institute{Leiden Observatory, Leiden University, Niels Bohrweg 2, 2333 CA, Leiden, the Netherlands\\
              \email{herle@strw.leidenuniv.nl}
         \and
              Institute for Theoretical Physics, Utrecht University, Princetonplein 5, 3584 CC, Utrecht, the Netherlands
            \and 
                Lorentz Institute for Theoretical Physics, Leiden University, PO box 9506, 2300 RA Leiden, the Netherlands\\
             }

  \abstract 
   {Intrinsic alignments are a major astrophysical contaminant for next generation large-sky surveys like \textit{Euclid} and LSST. Large hydrodynamic simulations are crucial for informing the alignment modelling for these surveys. We measure position-position and position-shape correlations of a Luminous Red Galaxy sample from the \pkg{FLAMINGO} suite of hydrodynamical simulations, measuring the alignment signal for more than 4.9 million galaxies at redshift 0. We jointly model the clustering and alignment correlations to provide the tightest constraints on the alignment amplitude to date from a hydrodynamic simulation. We find that both the Non-Linear Alignment (NLA) and the more complex Tidal Alignment Tidal Torquing (TATT) models provide good fits to the data. We compare the measured $A_1$ amplitude to observational data and find good agreement. We measure the dependence of the NLA and TATT free parameters on halo mass. We also introduce a mass-dependent TATT model, TATT-M, by finding empirical relations between the halo mass and the TATT parameters. This allows us to fit TATT with only one parameter, $A_1$, with $A_2/A_1$ being a constant and $A_{1\delta}/A_1$ being a function of halo mass. Using a Bayesian approach, we find that TATT-M is very strongly preferred by the data over NLA. Using the baryonic feedback variations of the \pkg{FLAMINGO} simulation suite, we test whether the TATT parameters are sensitive to feedback. Variations in AGN and supernova feedback do not significantly change the alignment amplitude beyond the change associated with the dependence of galaxy stellar mass on the strength of feedback. Our results inform the IA modelling for upcoming surveys by providing guidance on model choices, priors and sensitivities to feedback.}


   \maketitle

\section{Introduction}

The $\Lambda$CDM model for cosmology has been studied extensively in the last few decades, with analyses of the Cosmic Microwave Background (CMB) lending strong evidence in its favour \citep{Planck2020}. This model has two main ingredients: dark matter and dark energy. As the CMB is an early-Universe probe, its sensitivity comes from the distance to last scattering and thus does not probe the growth of the large-scale structure. It therefore has limited sensitivity to the dark energy equation of state compared to late-time probes. Instead, weak gravitational lensing of light coming to us from distant galaxies traces the intervening matter, and is sensitive to the matter distribution. Cosmic shear analyses measure the resulting correlation of distortions of galaxy images due to the effect of weak lensing, which remains the only way to directly probe the dark matter that makes up the cosmic web. Thus, cosmic shear has emerged as one of the primary probes of current galaxy surveys to explore dark energy.

Stage IV surveys like \textit{Euclid} \citep{Laureijs2011, Mellier2025} and the Legacy Survey of Space and Time \citep[LSST, ][]{LSST2009, Ivezic2019} will herald a new era of weak lensing analyses, with their unparalleled depth and area. This requires more sophisticated modelling of the shear signal than ever before, and systematics that were previously hidden in the noise can become a significant contributor to the signal. Since the effect of the foreground large-scale structure can be detected as a correlation of galaxy shapes in a weak lensing survey, the intrinsic alignment (IA) of galaxies is a major astrophysical systematic in these analyses \citep[see][for reviews]{Joachimi2015, Kirk2015, Kiessling2015, Lamman2024a, Chisari2025}.

IA induces a correlation in the shapes of galaxies that has been shown to contaminate weak lensing analyses \citep{Heavens2000, Catelan2001, Hirata2004} and bias cosmological parameters if modelled incorrectly \citep{Hirata2007, Samuroff2024}. The level of contamination from IA is about 10 per cent \citep{Chisari2015b}, which for surveys with precision requirements of 1 per cent represents a very important systematic. Therefore, many studies have attempted to detect and model the IA of galaxies in observational data  \citep[see ][for a recent example]{Navarro-Girones2025}, especially the alignment of red galaxies \citep{Okumura2009, Singh2015, Fortuna2021b, Zhou2023, Siegel2025a}. 

The most common method to mitigate IA contamination in weak lensing analyses is to include it in a joint modelling effort with the shear signal. Typical models used include the Non-Linear Alignment (NLA) model, with one free parameter, and the Tidal Alignment and Tidal Torquing (TATT) model, with two additional higher order parameters. It is very important to understand whether the existing IA modelling and mitigation strategies are sufficient for upcoming surveys. Moreover, sensible priors for the alignment parameters are required, because a totally agnostic approach would severely compromise the constraining power for the cosmological parameters. This requires simulations that are both large enough to be comparable to the survey-size of missions like \textit{Euclid} and LSST, and have sufficient resolution to resolve small-scale effects that influence the tidal field that causes the alignment of galaxies. This is vital as galaxies align with each other from the smallest scales, for example satellite-central alignments, to cosmological scales, like the alignment of galaxy clusters. Simulations thus need to not only have the resolution to resolve small scales, but also have the volume required to study large-scale alignments and produce enough haloes at the high-mass end. 

The dependence of IA on baryonic physics still represents an open question in the field \citep[e.g. ][]{Velliscig2015a, Tenneti2017, Soussana2020, Bilsborrow2025}. None of the alignment models in the literature take feedback into account. As observational data improves, so does our ability to push to smaller scales, which are more strongly influenced by non-linear effects. How feedback affects alignments is difficult to test with simulations, as different simulations incorporate different sub-grid physics, thereby making a one-to-one comparison difficult \citep[see][ for an attempt to do so]{vanHeukelum2025b}. 

The need to understand the limitations of our IA modelling strategies requires a new era of cosmological simulations. Several studies have used both gravity-only simulations and hydrodynamic simulations to study IA. For a compilation of such studies, see Table 1 of \citet{Chisari2025}. The \textit{Euclid} Flagship simulation \citep{Castander2025} is one such effort to help understand \textit{Euclid} observations. It has a box-size of 3.6 $h^{-1}$Gpc and a particle mass of $10^9 \ h^{-1} \ \mathrm{M}_{\odot}$, and has been used to model IA for the \textit{Euclid} analyses \citep[][Euclid Collaboration: Navarro-Girone\'s et al. in prep.]{Hoffmann2026, Paviot2026}. Flagship is a gravity-only simulation, and galaxies are placed in dark matter haloes using a halo occupation distribution (HOD) and a halo abundance matching approach \citep{Carretero2015, Castander2025}. The \pkg{FLAMINGO} suite of simulations \citep{Schaye2023, Kugel2023}, which we used in this work, provides a comparable box size and resolution to that of Flagship, but has the added advantage of being a fully hydro-dynamic simulation. This allows us to study alignments without making the assumptions on the galaxy-halo connection that are central to the HOD method. Moreover, \pkg{FLAMINGO} has feedback variations that are useful for testing the effect on the alignment signal. The galaxy sample we use has halo masses similar to those that host Luminous Red Galaxies (LRGs) which can be used to compare with LRG results from observations, and also to inform priors on future LRG studies.

In this work, we use the \pkg{FLAMINGO} simulation suite to study IA. In Section \ref{sect:formalism}, we describe the formalism and modelling of the IA signal. In Section \ref{sect:flamingo}, we describe the \pkg{FLAMINGO} suite of simulations. In Section \ref{sect:comparison} we compare our best-fitting parameters with observational studies. In Section \ref{sect:mass-dependence}, we explore the dependence of the non-linear galaxy bias and alignment terms under both NLA and TATT, and in Section \ref{sect:feedback}, we explore the effect of feedback on the TATT parameters. 

\section{Formalism}
\label{sect:formalism}

In this section we describe the mathematical framework for the measurements described in this work. We focused on the 2-point clustering and alignment signal, which are the auto-correlations of positions and the cross-correlations of shapes and positions of objects. In the notation that follows, these objects are galaxies and are denoted by the subscript `g'. The subscript '+' denotes the shapes of galaxies.

The matter power spectrum encodes information about the density field, and a Fourier transform of the matter power spectra results in the 3D correlation functions

\begin{equation}
    \xi_{\mathrm{AB}}(r, z) = \int\frac{\mathrm{d}^3 k} {(2\pi)^3}\mathrm{P_{AB}}(k, z) e^{i(k.r)}.
\end{equation}

\noindent where $k$ is the wavenumber, $z$ is the redshift, $r$ is the 3D separation, and A and B correspond to two tracers. For example, if A and B were both galaxy position samples, then this would give the position-position correlation. In lensing observations, we are limited to projected quantities, so we calculate the correlation functions by integrating the 3D ones along the line-of-sight separation $\Pi$:
\begin{equation}
\label{eqn:wab}
    w_{\mathrm{AB}}(r_{\mathrm{p}}) = \sum^{\Pi_{\mathrm{max}}}_{0} \Delta \Pi \,\xi_{\mathrm{AB}},
\end{equation}

\noindent where $r_{\mathrm{p}}$ is the 2D separation, $\Pi_{\mathrm{max}}$ is the maximum line-of-sight separation integrated over, which can be used for the position-position (`gg') and position-shape (`g+') correlation functions \citep{Blazek2011, Singh2016}

\begin{align}
\label{eqn:wgg}
w_{\mathrm{gg}}(r_{\mathrm{p}}) &= \frac{1}{\pi^2} 
\int_0^{\infty} \mathrm{d} k_z 
\int_0^{\infty} \mathrm{d} k_{\perp} \, 
\frac{k_{\perp}}{k_z} \,\nonumber\\
&\quad \times P_{\mathrm{gg}}(\vec{k}, z_s)  \sin(k_z \Pi_{\mathrm{max}}) \, J_0(k_{\perp} r_{\mathrm{p}}),
\end{align}

\noindent and 

\begin{align}
\label{eqn:wgp}
w_{\mathrm{g+}}(r_{\mathrm{p}}) &= \frac{1}{\pi^2}
\int_0^{\infty} \mathrm{d} k_z 
\int_0^{\infty} \mathrm{d} k_{\perp} \,
\frac{k_{\perp}^3}{k_{\perp}^2 + k_z^2} \, \nonumber\\
&\quad \times P_{\mathrm{gI}}(\vec{k}, z_s) \sin(k_z \Pi_{\mathrm{max}}) \, J_2(k_{\perp} r_{\mathrm{p}}),
\end{align}

\noindent where $J_0$ and $J_2$ are cylindrical Bessel functions of the first kind, of order 0 and 2 respectively, $k_{\perp}$ is the perpendicular wavenumber, and $z_s$ is the redshift of the simulation snapshot. $P_{\mathrm{gg}}$ denotes the auto-correlation of galaxy positions, and $P_{\mathrm{gI}}$ quantifies the correlation between the galaxy positions and the intrinsic ellipticities. The expressions for these power spectra can be derived from the matter and galaxy fields $\delta_{\mathrm{m}}$ and $\delta_{\mathrm{g}}$, related by the galaxy bias term, $b_1$, as

\begin{equation}
\label{eqn:lin_bias}
    \delta_{\mathrm{g}} = b_1 \delta_{\mathrm{m}},
\end{equation}

\noindent which assumes the linear bias model that is applicable at large scales \citep{Kaiser1984}. To extend this model to smaller scales, where non-linear effects play a more dominant role, we can express the galaxy field as \citep{McDonald2006, Baldauf2010, Saito2014}:

\begin{equation}
    \delta_{\mathrm{g}} = b_1 \delta_{\mathrm{m}} + \frac{1}{2}b_2(\delta_{\mathrm{m}}^2- \langle \delta_{\mathrm{m}}^2 \rangle)
    + \frac{1}{2}b_{s^2}(s^2- \langle s^2 \rangle) + b_{3\mathrm{nl}}\psi,
\end{equation}

\noindent where $s$ is the tidal field, and $s^2 = s_{ij}s^{ij}$, using the Einstein summation convention, $\psi$ is the sum of the third-order non-local terms with the same scaling, $b_2$ the local quadratic bias, $b_{s^2}$ the tidal quadratic bias and $b_{3\mathrm{nl}}$ is the third-order non-local bias. The galaxy-galaxy power spectrum can then be written as (from Eqn. 38 of \citealt{Krause2021})

\begin{align}
P_{\mathrm{gg}}(k) &= b_1^2P_{\delta \delta}(k) + b_1 b_2P_{b_1 b_2}(k) \nonumber\\
&\quad + b_1 b_{s^2}P_{b_1 s^2}(k) + b_1 b_{3\mathrm{nl}} P_{b_1 b_{3\mathrm{nl}}}(k) \nonumber\\
&\quad + \tfrac{1}{4}b_2^2P_{b_2 b_2}(k)
 + \tfrac{1}{2}b_2b_{s^2}P_{b_2 s^2}(k)
 + \tfrac{1}{4}b_s^2P_{s^2 s^2}(k),
\end{align}

\noindent where, $P_{\delta \delta}$ is the non-linear matter power spectrum and the power spectrum kernels $P_{b_1}$, $P_{b_2}$, $P_{b_1s^2}$ etc. are defined in \citet{Saito2014}. We also invoke the co-evolution relations $b_{s^2} = -\frac{4}{7} (b_1 -1) $ and $b_{3\mathrm{nl}} = b_1 -1$ to reduce our parameter space \citep{Saito2014}. Finally, $P_{\mathrm{g+}}$ can be expressed as 

\begin{equation}
    P_{\mathrm{g+}}(k, z) = b_1 P_{\delta+}(k, z),
\end{equation}

\noindent where $P_{\delta+}$ is the matter-intrinsic power spectrum.

We compare the IA power spectrum with two models, the NLA and the TATT models. The position-shape correlation function $w_{\mathrm{g+}}$ constrains the product $A_1 b_1$, while joint modelling with the clustering correlation function $w_{\mathrm{gg}}$, which constrains $b_1$, allows us to break the degeneracy between these parameters. In this work, even though we use the non-linear galaxy bias to model the clustering, the IA models we use inherently assume only a linear bias. We do not expect this to affect our results, since the bias terms are almost entirely constrained by the clustering signal due to its higher statistical power, and the alignment signal provides little to no constraining power for the bias terms. Moreover, including non-linear bias terms in the alignment modelling would introduce higher order terms that we have implicitly ignored, as they should be lower in magnitude than the terms considered. We achieved reasonable fits even without these higher order terms.

\subsection{NLA}

The linear alignment model \citep{Catelan2001, Hirata2004} assumes that the alignments of galaxies are imprinted at the time of galaxy formation by the initial tidal field of the galaxy's environment. Since this model stems from linear theory alone, it performs well only on  large scales ($>$ 100 Mpc). This model originally uses the linear matter power spectrum, but using the non-linear matter power spectrum instead, as proposed by \citet{Hirata2004} and implemented by \citet{Bridle2007}, attempts to extend this model to quasi-linear scales. This is known as NLA model. Following \citet{Hirata2004}, the intrinsic ellipticity of an object can be written as:

\begin{equation}
    \epsilon = - \frac{\bar{C}_1}{4\pi G}(\Delta_x^2 - \Delta_y^2, 2\Delta_x \Delta_y)\textbf{\cal{S}}[\psi_P],
\end{equation}

\noindent where $G$ is the Newtonian gravitational constant, $x$ and $y$ are the Cartesian coordinates in the plane of the sky, $\textbf{\cal{S}}$ is a smoothing filter that cuts off fluctuations on galactic scales for $\psi_P$, which is the Newtonian potential at the time of galaxy formation. $\Delta$ is the comoving derivative, and $\bar{C_1} = 5 \times 10^{-14} \mathrm{M}_{\odot}^{-1}h^{-2} \mathrm{Mpc^3}$ is a normalisation constant set by \citet{Brown2002} for low-redshift IA measurements in SuperCOSMOS \citep{Hambly2001}. Following this model,

\begin{equation}
    P_{\delta +}(k, z) = C_1(z) P_{\delta \delta}(k, z),
\end{equation}

\noindent with 
\begin{equation}
\label{eqn:a1_norm}
    C_1(z) = - A_1 \bar{C_1} \rho_{\mathrm{crit}} \Omega_{\mathrm{m}}D(z)^{-1},
\end{equation}

\noindent where $A_1$ is the IA amplitude, $\rho_{\mathrm{crit}}$ is the critical density, $\Omega_{\mathrm{m}}$ is the fractional matter density and $D(z)$ is the linear growth factor normalised such that is 1 at $z = 0$. Following the linear alignment model of \citet{Hirata2004}, the alignments are assumed to be established at the time of formation of the galaxy or as a power-law of redshift.

The NLA model has been shown to perform well on intermediate scales in both observational and simulation-based studies \citep{Fortuna2025, Paviot2026}. It involves fitting two parameters: the galaxy bias $b_1$ and the intrinsic alignment amplitude $A_1$.

\subsection{TATT}

The TATT model was proposed by \citet{Blazek2019} and incorporates linear and quadratic terms in the alignment signal. Below we review the formalism of this model analogously to our treatment of the NLA model. An expansion of the density field up to second-order gives the intrinsic shear field \citep[Eqn. 8 of][]{Blazek2019}:

\begin{equation}
    \gamma_{ij}(\textbf{x}) = C_1 s_{ij} +C_2 \big ( s_{ik}s_{kj} -\frac{1}{3}\delta_{ij}s^2 \big) + C_{1\delta}(\delta s_{ij}).
\end{equation}

\noindent where the fields are evaluated at \textbf{x}, and summation over repeated indices is implied. $C_2$ and $C_{1\delta}$ are additional bias terms that capture the strength of the higher order contributions. The observable ellipticity components in projection then become \citep[Eqn. 13 of][]{Blazek2019}

\begin{equation}
    (\gamma_+, \gamma_{\times}) = (C_1 + C_{1\delta})(s_{xx} - s_{yy}, 2s_{xy}) + C_2(s_{xk}s_{xk}-s_{yk}s_{yk}, 2s_{xk}s_{yk}).
\end{equation}

\noindent $P_{\delta +}$ can then be written as \citep[Eqn. 21 of ][]{Secco2022}:

\begin{equation}
\label{eqn:tatt_pgi}
    P_{\delta +}(k, z) = C_1(z)P_{\delta \delta}(k, z) + C_{1\delta}(z)P_{0|0E}(k, z) + C_2(z)P_{0|0E2}(k, z).
\end{equation}

$P_{0|0E}$ and $P_{0|0E2}$ are the relevant power spectra, and the notation is consistent with that used by \citet{Blazek2011}, and we refer the reader to that work for the exact forms of these power spectra and a more detailed derivation.The normalisation of $A_1$ follows the NLA prescription given in Eqn. \ref{eqn:a1_norm}. The additional normalization terms required for the TATT model can be written as \citep[Eqns. 41 and 42 of][]{Blazek2011}

\begin{align}
    C_{1\delta}(z) &= -A_{1\delta}(z)(\bar{C}_1\rho_{\mathrm{crit}})\Omega_{\mathrm{m}}D(z)^{-1},\\
    C_2(z) &= A_2(z) \bigg( \frac{5\bar{C}_1 \rho_{\mathrm{crit}}}{\Omega_{\mathrm{m, fid}}} \bigg)\Omega_{\mathrm{m}}^2D(z)^{-2}.
\end{align}

The NLA model can be interpreted as a subset of the TATT model, with only $A_1$ set to be non-zero.\footnote{Note that the $A_1$ in NLA and TATT are in principle the same quantity. In practice, however, depending on the strength of the higher order alignment terms, the value of $A_1$ in NLA and TATT may be different.} Within the TATT model for alignment, the signal is additionally represented by $A_2$ and $A_{1\delta}$, which allows the extension of this model to more non-linear scales than viable with NLA. More complex models do exist in the literature, like Effective Field Theory \citep[EFT, ][]{Bakx2023, Chen2024} and HYMALAIA \citep{Maion2024}. Since we are interested in testing the validity of NLA and TATT on the \pkg{FLAMINGO} data, we limit the model complexity in this work to just these two models. 

\subsection{Estimators}
In order to calculate the correlation functions, we employ the generalised Landy-Szalay correlation function estimators \citep{Landy1993}, which are defined as:

\begin{equation}
    \xi_{\mathrm{gg}} = \frac{(S - R_S)(D - R_D)}{R_S R_D} = \frac{SD - R_SD - SR_D + R_S R_D}{R_S R_D},
\end{equation}

\noindent and 

\begin{equation}
    \xi_{\mathrm{g}+} = \frac{S_+D - S_+R_D}{R_S R_D},
\end{equation}

\noindent where $S$ is the shape sample and $D$ is the position sample. $R_S$ and $R_D$ are the shape and position random samples respectively. For all measurements in this work, we employ a random catalogue that is ten times larger than the dataset. We verified convergence by comparing with a catalogue twenty times larger and found that the increase does not change our results. The terms containing shears are calculated using

\begin{equation}
    S_+X = \sum_{i\in S, j\in X} \gamma_+^{(i)}\langle j | i \rangle.
\end{equation}
\noindent where $X$ is either $D$ or $R_D$, $\gamma_+$ measures the components of the shear along the line joining the pair of galaxies, and  $\langle j | i \rangle$ implies galaxy shape $j$ with respect to separation vector towards galaxy $i$.

All correlation functions presented in this work were computed using \pkg{IACorr},\footnote{https://github.com/elizabethjg/IACorr} which is built upon \pkg{TreeCorr} \citep{Jarvis2004}. We used the delete one jack-knife (JK) method to estimate covariance matrices \citep[for more details, see][]{Norberg2009}.\footnote{We employed a version of this code in which the projection of the 3D correlation function along the $\Pi$ direction is performed via direct summation, rather than trapezoidal integration, over only one $\Pi$ bin. This has been validated against an independent implementation also based on \pkg{TreeCorr} \citep{Jarvis2004}, called \pkg{MANTIS}, as well as two independent codes, \pkg{MeasureIA} and \pkg{HaloTools}, that do not rely on \pkg{TreeCorr}.} The covariance matrix is given by

\begin{equation}
\mathrm{Cov}_{\mathrm{JK}} =
\frac{N_{\mathrm{JK}}-1}{N_{\mathrm{JK}}}
\sum_{i=1}^{N_{\mathrm{JK}}}
\bigl(
\mathbf{w}_{\mathrm{ab},i}-\bar{\mathbf{w}}_{\mathrm{ab}}
\bigr)
\bigl(
\mathbf{w}_{\mathrm{ab},i}-\bar{\mathbf{w}}_{\mathrm{ab}}
\bigr)^{\mathrm{T}},
\end{equation}

\noindent where $N_{\mathrm{JK}}$ is the number of jack-knife patches, $\mathbf{w}_{\mathrm{ab},i}$ is the correlation function after removing the signal of the $i$th JK region and $\bar{\mathbf{w}}_{\mathrm{ab}}$ is the mean of all the JK regions. Unless otherwise mentioned, $N_{\mathrm{JK}} = 125$ for all the error bars in this paper. 

We jointly modelled $w_{\mathrm{gg}}$ and $w_{\mathrm{g+}}$. To model our data vectors, we developed a Python package called \pkg{IATheory} \citep{Navarro-Girones2025} based on the formalism described in this section, which will be publicly available soon. This code is based on {\tt pyccl} \citep{Chisari2019}, and uses \pkg{scipy} \citep{scipy} for the Bessel function and integration step in Eqns. \ref{eqn:wgg} and \ref{eqn:wgp}. For fitting the model, it implements the {\tt nautilus} sampler \citep{nautilus}, which is a nested sampling code that uses machine learning to explore the parameter space efficiently. This code has been validated against three other codes developed independently to model $w_{\mathrm{gg}}$ and $w_{\mathrm{g+}}$, and we found consistent results when using the same data vector (see Acknowledgements). We also obtain consistent posteriors when we used the Markov chain Monte Carlo (MCMC) sampling within \pkg{IATheory} with {\tt emcee} \citep{emcee}, also available in \pkg{IATheory}.
\begin{table}
\caption{Simulation details of the feedback variations used in this work.}
\label{table:sim_details}
\centering
\begin{tabular}{ccccrS[table-format=3.2]} 
\hline
\noalign{\smallskip}
Identifier & $\Delta m_*$ & $\Delta f_{\mathrm{gas}}$ & AGN & $N_\mathrm{b}$ \\
 & ($\sigma$) & ($\sigma$) &  &  \\
 \noalign{\smallskip}
  \hline
  
  L1\_m9 & 0 & 0 & thermal & $1800^3$ \\
  fgas+2$\sigma$ & 0 & +2 & thermal & $1800^3$ \\
  fgas-2$\sigma$ & 0 & -2 & thermal & $1800^3$ \\
  fgas-4$\sigma$ & 0 & -4 & thermal & $1800^3$ \\
  fgas-8$\sigma$ & 0 & -8 & thermal & $1800^3$ \\
  M*-$\sigma$ & -1 & 0 & thermal & $1800^3$  \\
  Jet & 0 & 0 & jets & $1800^3$ \\
  Jet fgas-4$\sigma$ & 0 & -4 & jets & $1800^3$ \\
  L2p8\_m9 & 0 & 0 & thermal & $5040^3$ \\
  \hline
\end{tabular}
\tablefoot{All runs have a mean Cold Dark Matter (CDM) particle mass $\mathrm{m}_{\mathrm{CDM}} = 5.65 \times 10^9\mathrm{M}_{\odot}$. The L2p8\_m9 run has a box size of 2.8 Gpc while the rest have a box size of 1 Gpc. Identifier refers to the simulation run, $\Delta m_*$ is the number of standard deviations by which the observed stellar masses are shifted before calibration, $\Delta f_{\mathrm{gas}}$ is the number of standard deviations the observed cluster gas fractions are shifted before calibration, AGN is the form of AGN feedback implemented and $N_\mathrm{b}$ is the number of baryonic particles. All these runs assume the fiducial cosmology with $h$ = 0.681, $\Omega_{\mathrm{m}} = 0.306$, $\Omega_{\Lambda} = 0.694$ , $\Omega_\mathrm{b} = 0.0486$,  $\Sigma \mathrm{m_\nu} \mathrm{c}^2 = 0.06$eV, $\mathrm{A}_\mathrm{s} = 2.099 \times 10^{-9}$, $\mathrm{n}_{\mathrm{s}} = 0.967$, $\sigma_8 = 0.807$, $\Omega_{\nu} = 1.39\times 10^{-3}$}  
\end{table}

\begin{figure*}
    \centering
    \includegraphics[width=\textwidth]{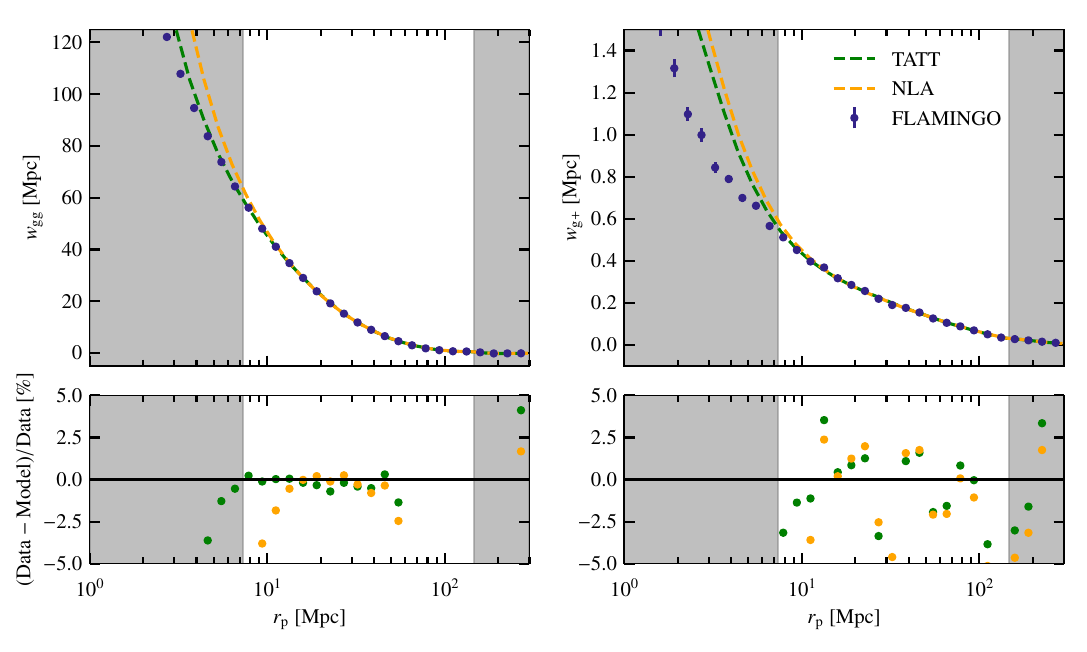}
    \caption{The projected position-position, $w_{\mathrm{gg}}$ (left) and position-shape, $w_{\mathrm{g+}}$ (right), correlation functions for all galaxies with more than 300 stellar particles in the 2.8 Gpc$^3$ box, as a function of the projected separation, $r_{\mathrm{p}}$. The sample contains $4\,941\,492$ galaxies with stellar masses ranging from $10^{11.28}$ to $10^{13.68} \ \mathrm{M}_{\odot}$ and halo masses ($\mathrm{M}_{200\mathrm{mean}}$) ranging from $10^{12.2}$ to $10^{15.78} \ \mathrm{M}_{\odot}$. Overplotted is the best-fitting joint clustering and NLA or TATT model with the residuals in the lower panel. The gray regions show the scales excluded from the fitting. Considering scales between 5 and 100 Mpc/$h$, TATT achieves residuals within 4 per cent.}
    \label{fig:2p8_fits}
\end{figure*}

\begin{figure}
    \centering
    \includegraphics[width=\columnwidth]{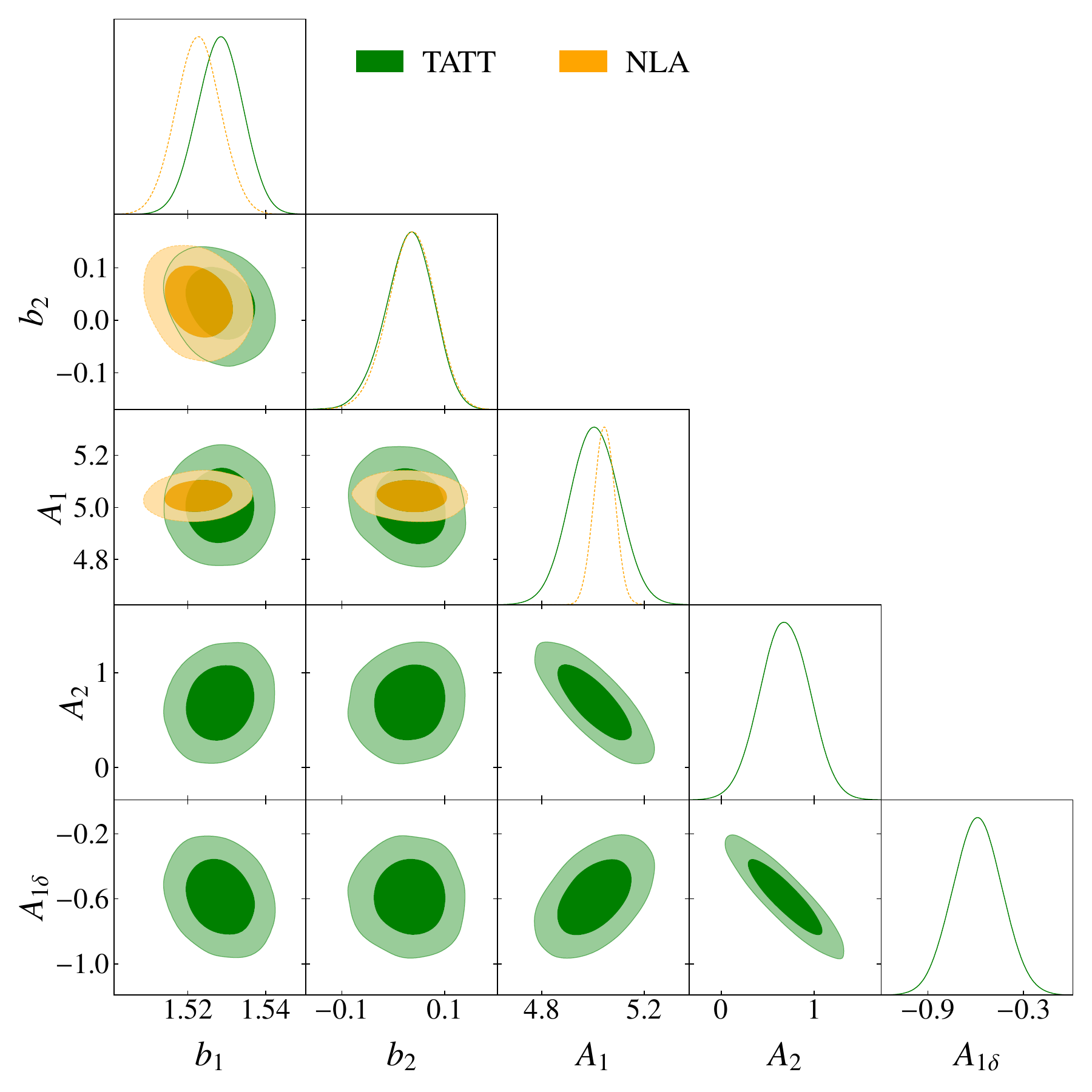}
    \caption{The best-fitting TATT and NLA model parameters for the 2.8 Gpc sample. Only scales within 5-100 Mpc/$h$ are modelled. The posteriors of NLA and TATT are consistent with each other.}
    \label{fig:2p8_contour}
\end{figure}

\section{\pkg{FLAMINGO} simulations}
\label{sect:flamingo}

Throughout this paper, we used the output of \pkg{FLAMINGO}, which is a Virgo consortium project presented in \citet{Schaye2023} and \citet{Kugel2023}. \pkg{FLAMINGO} is a large suite of cosmological structure formation simulations that encompass variations in cosmology, baryonic feedback and numerical resolution. Particularly useful for this work are the large box sizes, with the largest box with hydro-dynamics being 2.8 Gpc on a side, while the runs varying the feedback and cosmology are 1 Gpc to a side. These produce a large number of haloes that allow us to measure the alignment signal with unprecedented precision. 

The simulations were performed with the SWIFT hydrodynamics code \citep{Schaller2024}. The simulations include neutrinos \citep{Elbers2021}, radiative processes \citep{Ploeckinger2020}, star formation \citep{Schaye2008}, stellar mass loss and enrichment \citep{Wiersma2009, Schaye2015}, kinetic stellar feedback \citep{Chaikin2023}, black hole growth and AGN feedback \citep{Booth2009, Husko2022}.

The sub-grid prescriptions of stellar and AGN feedback were calibrated to the observed low redshift cluster gas fractions and galaxy stellar mass function, as described by \citet{Kugel2023}. The 1 Gpc$^3$ runs have identical initial conditions but different particle mass resolutions. The effect of resolution on halo shapes and orientations was explored by \citet{Herle2025}, and in this work we use only the runs with the fiducial resolution of $\mathrm{m}_{\mathrm{CDM}} = 5.65 \times 10^9 \ \mathrm{M}_{\odot}$. The different feedback modes implemented varied the AGN feedback mechanism between thermal and jet feedback. As the observations to which the models were calibrated have uncertainties, several feedback variations were created by calibrating the sub-grid parameters to gas fractions or stellar mass functions that are shifted away from their fiducial values, as described by \citet{Kugel2023}. Each variation was defined with respect to the observable it was calibrated to, that is the number of standard deviations by which the observed stellar masses and cluster gas fractions were shifted prior to calibrating the sub-grid parameters. All the simulation runs used in this work are summarised in Table \ref{table:sim_details}.

Halo finding is a crucial step in analysing cosmological simulation data, and the choice of halo finder has been shown to influence results strongly \citep{Knebe2013a, Knebe2013b,  Moreno2025}. A 3D Friends-of-Friends (FoF) algorithm with linking length $l = 0.2$ times the mean CDM interparticle separation was first used to group particles using only the dark matter particles. Gas and stellar particles are then attached to the nearest CDM particle. \pkg{HBT-HERONS} \citep[Hierarchical Bound Tracing - Hydro-Enabled Retrieval of Objects in Numerical Simulations, ][]{Moreno2025}, which is a modified version of HBT+ \citep{Han2018}, then tracks haloes across timesteps to produce a subhalo catalogue. We use \pkg{HBT-HERONS} to produce subhalo catalogues, since it performs better than other subhalo finders \citep{Moreno2025}. We used catalogs of simulation quantities calculated by Spherical Overdensity and Aperture Processor (SOAP) \citep{McGibbon2025}.

\subsection{Inertia tensors and shapes}

To estimate the shape of a given halo, we calculated its inertia tensor. Haloes were modelled as 3D ellipsoids whose axes point in the direction of the eigenvectors of the Simple Inertia Tensor (SIT) defined as:

\begin{equation}
\label{eqn:sit}
    I_{ij} = \frac{1}{M}\sum_n m_{(n)}x_{i}^{(n)}x_{j}^{(n)},
\end{equation}
where $i, j = {1, 2, 3}$ correspond to the three axes of the simulation box, $m_n$ is the mass of the $n$th particle, $x_{i, j}^{(n)}$ are the positions of the $n$th particle in the $i$ or $j$ direction and $M$ is the total mass of the object. These were implemented in SOAP \citep{McGibbon2025} and were calculated for all bound particles within the half-mass radius of the objects. There exist four principal formulations of the inertia tensor, each optimized for specific applications. We use the SIT scheme because this choice maximises the signal-to-noise ratio of the alignment signal. We compare how these different definitions influence our results in Appendix \ref{appendix:inertia_tensors}, finding that changing the definition of the inertia tensor introduces only a shift in the amplitude of the alignment signal. 

The eigenvalues of the inertia tensor in Eqn. \ref{eqn:sit} are denoted $\lambda_i$, where $i = {1, 2, 3}$ correspond to the three axes of the ellipsoid. The lengths of the three axes of this ellipsoid, the major, intermediate and minor axes, are given by $a = \sqrt{\lambda_1}$, $b = \sqrt{\lambda_2}$ and $c = \sqrt{\lambda_3}$ respectively. We further define the axis ratio $q = b/a$. The eigenvectors of the inertia tensor, $\hat{e}_i$, encode the orientation of the object.

\subsection{Correlation functions in \pkg{FLAMINGO}}

For our galaxy sample, we selected all galaxies from subhaloes with a bound dark matter mass $>10^{12} \ \mathrm{M}_{\odot}$ with more than 300 stellar particles, corresponding to a stellar mass of $\approx 2.22 \times10^{11}\ \mathrm{M}_{\odot}$ in the fiducial feedback variation. With these cuts, we ensured that our dark matter and stellar shapes are both well resolved, and that we could compare the halo and galaxy alignment signals which we will do in a follow-up analysis. In the rest of this paper, we refer to the sample created after applying the cuts mentioned above when we refer to our galaxy sample. The range of halo and stellar masses of the resulting sample is similar to that of Luminous Red Galaxies (LRGs), and looking at the color-magnitude diagram (\textit{u -r}), we found that our sample is very red.

In Fig. \ref{fig:2p8_fits}, we show $w_{\mathrm{gg}}$ and $w_{\mathrm{g+}}$ for our galaxy sample from the 2.8 Gpc simulation run, with $\Pi_{\mathrm{max}}=100$ Mpc/$h$. The sample contains $4\,941\,492$ galaxies with stellar masses ranging from $10^{11.28}$ to $10^{13.68} \ \mathrm{M}_{\odot}$ with a median mass of $3.4 \times 10^{11} \ \mathrm{M}_{\odot}$. The host haloes of this sample have halo masses ($\mathrm{M}_{200\mathrm{mean}}$) ranging from $10^{12.2}$ to $10^{15.78} \ \mathrm{M}_{\odot}$, with a median of $1.61 \times 10^{13} \ \mathrm{M}_{\odot}$. Overplotted are our best-fitting models for NLA and TATT, fit jointly with the clustering, and in the bottom panel we show the residuals between the models and data. We define our residuals as (Data - Model)/Data expressed as a percentage. The grey regions in the plot depict the scales excluded from the fitting, with scale cuts of $r_{\mathrm{min}} = 5$ Mpc/$h$ and $r_{\mathrm{max}} = 100$ Mpc/$h$. We achieve a model error of only a few per cent using non-linear bias to model the clustering signal and NLA or TATT to model the alignment signal. The TATT model has smaller residuals at smaller scales than the NLA model, consistent with the picture that TATT captures more higher order information. Using the reduced chi-squared ($\chi^2_{\mathrm{red}}$) as a metric for goodness of fit, we see that TATT fits the data better for a minimum scale ($r_{\mathrm{min}}$) of 5 Mpc/$h$. To achieve approximately the same $\chi^2_{\mathrm{red}}$ with NLA, we needed to increase $r_{\mathrm{min}}$ to 10 Mpc/$h$. If LRGs contribute to most of the IA contamination in lensing, then the NLA model is sufficient down to scales of 10 Mpc/$h$ for the analysis of next generation survey data since the error bars from our galaxy sample are smaller than the statistical errors that will be obtained in such surveys, and in order to include data down to 5 Mpc/$h$, TATT would be required.

The posteriors\footnote{Whenever we model a data vector, we use the {\tt nautilus} sampler, with the following settings: $N_{\mathrm{live}} =  10 \ 000$, number of networks = 16, and {\tt discard\_exploration=True}. These settings ensure that the final posteriros are well sampled.} for these fits are shown in Fig. \ref{fig:2p8_contour}. NLA and TATT are consistent with each other when the same scale cuts are used, and since there are fewer parameters for the NLA model, the constraints on $b_1$, $b_2$ and $A_1$ are tighter. The TATT parameters $A_2$ and $A_{1\delta}$ are correlated with each other and also with $A_1$. This indicates that these parameters can be related to each other, potentially reducing the parameter space for a TATT model fit. This is explored in Section \ref{sect:mass-dependence}.

\section{Comparison with observations}
\label{sect:comparison}

\begin{figure}
    \centering
    \includegraphics[scale=0.35]{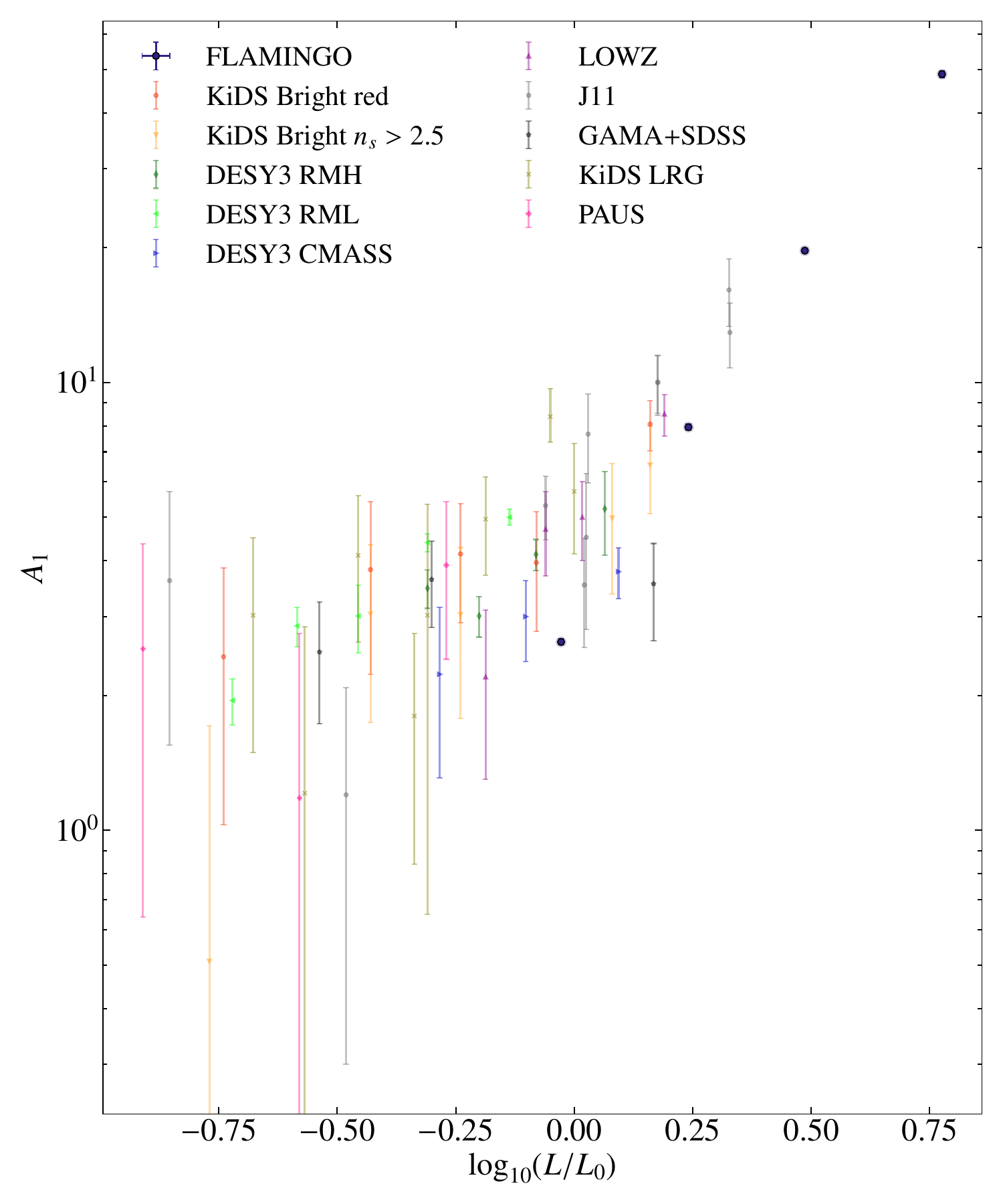}
    \caption{The alignment amplitude with NLA for different \textit{r}-band luminosity bins, $L_0$ is the pivot luminosity, with a value of $4.6\times 10^{10} \ h^{-2} \ \mathrm{L}_{\odot}$. The galaxy sample from the 2.8 Gpc$^3$ run at $z = 0$ was first binned in stellar mass. The mean halo mass for each bin was then converted to an \textit{r}-band luminosity using a fit to the data in Table 4 of \citet{Mandelbaum2006a}. The \pkg{FLAMINGO} values, shown in deep blue, agree reasonably well with observational studies of LRGs, and extends the range to higher luminosities. Error bars show the 1$\sigma$ error on the best-fitting $A_1$. Error bars are omitted for the x-axis.}
    \label{fig:aia_vs_L}
\end{figure}

We compare the best-fitting $A_1$ with NLA for our galaxy sample with various observational studies on LRGs. We plot our measurements from the \pkg{FLAMINGO} sample with several results from the literature in Fig. \ref{fig:aia_vs_L}. We binned our galaxy sample in stellar mass and measured both $w_{\mathrm{gg}}$ and $w_{\mathrm{g+}}$ with $\Pi_{\mathrm{max}}=20$ Mpc/$h$ to obtain a high signal-to-noise ratio, and we jointly modelled them using the NLA alignment model. We convert the mean halo mass (($\mathrm{M}_{200\mathrm{mean}}$)) of each bin to an $r$-band luminosity using a power-law fit to data from Table 4 of \citet{Mandelbaum2006a}, allowing us to bypass any complications related to dust corrections in luminosities.

We briefly review the works considered in Fig. \ref{fig:aia_vs_L} here. (1) \citet{Joachimi2011} used the MegaZ-LRG sample from SDSS to measure alignments for galaxies with $z \lesssim 0.7$ (`J11'). (2) \citet{Singh2015} considered the SDSS-III BOSS LOWZ sample, in the redshift range $0.16 < z < 0.36$, with $\langle M_g - M_i \rangle = 1.18$ (`LOWZ'). (3) \citet{Johnston2019} used a KiDS+GAMA ($\langle z \rangle \approx 0.23$) and SDSS ($\langle z \rangle \approx$ 0.11, their Table 2) dataset to constrain the alignment amplitude of galaxies with $g - r > 0.66$ (`GAMA+SDSS'). (4) \citet{Fortuna2021b} used the KiDS-1000 dataset with LRGs in the redshift range $0.2 < z < 0.8$ to constrain the alignment amplitude (`KiDS LRG'), as well as combined data from previous works \citep{Joachimi2011, Singh2015, Johnston2019} to constrain double power-law relation between luminosity and alignment amplitude. (5) \citet{Samuroff2023} used measurements from photometric red sequence (redMaGiC) galaxies from DES with $\langle z \rangle = 0.78 $ (`DESY3 RMH'), $\langle z \rangle = 0.46$ (`DESY3 RML') and also a SDSS-III BOSS CMASS sample at $z \approx 0.5$ (`DESY3 CMASS'). (6) \citet{Georgiou2025} looked at the KiDS Bright sample, by selecting galaxies with an $r$-band magnitude $r < 20$ (`KiDS Bright red' in Fig. \ref{fig:aia_vs_L}) and then those with a Sérsic index more than 2.5 (`KiDS Bright $n_s > 2.5$'). The galaxies had redshifts $0.1 < z < 0.5$ (their fig. 2). (7) \citet{Navarro-Girones2025} extended the analyses to fainter galaxy luminosities using data from the PAU survey with $0.1 < z < 1$ and a magnitude limit $i_{\mathrm{AB}} < 22$ (`PAUS').

The magnitudes used in observations need to be corrected to $z = 0$ (\textit{k}-correction), and the evolution of the stellar population with redshift can be accounted for (\textit{e}-correction). Of the studies we include here, \citet{Joachimi2011}, \citet{Singh2015}, \citet{Fortuna2021b} and \citet{Samuroff2023} use \textit{k+e} corrected luminosities. \citet{Johnston2019} and \citet{Georgiou2025} do not mention whether their luminosities are \textit{k+e} corrected. \citet{Navarro-Girones2025} only apply \textit{k}-corrections to their data. These differences are expected to be small and we ignore them in this comparison.

Comparing our results (`FLAMINGO') with the observational studies, we find reasonable agreement. Given the large box size, we have a large number of bright galaxies, thereby extending the alignment amplitudes to the bright-end. Due to the resolution of the simulation, the number of faint galaxies is too low to capture the plateauing of the stellar-luminosity relation at the faint-end as reported by \citet{Fortuna2021b} and \citet{Navarro-Girones2025}.

An exact quantitative comparison of the alignment amplitudes in these works is difficult because of the differences in each study. Although they have all looked at the alignment of LRGs \citep[and have been used together for a joint fit on the power-law relation for NLA-M previously in ][]{Fortuna2021b}, the color cuts used to select the LRGs are different. Moreover, the redshift range for each of these works is different, although there has been no observed detection of redshift evolution in the literature thus far \citep[][for example]{Navarro-Girones2025}. Another issue arises from the different shape measurement techniques in each survey. Our shape measurement is consistent with the formalisms used in the literature, but different physical scales may be probed. Finally, we note that the \textit{r}-band luminosities can be different in different surveys due to the fact that they have different filter transfer functions, but this effect is expected to be very minor.

Even though the simulations were not calibrated to reproduce alignments in any way, we are able to recover values that are comparable to observations and trends with e.g. mass are captured well by the simulations. In order to make more accurate survey-specific comparisons, we would need to make mock images on the lightcones from the simulation for all these different surveys to account for different sensitivities, depth, seeing etc. In the future, with \textit{Euclid}, we would be able to probe the entire range of luminosities with high signal-to-noise with a single experiment.

\begin{figure}
    \centering
    \includegraphics[scale=0.5]{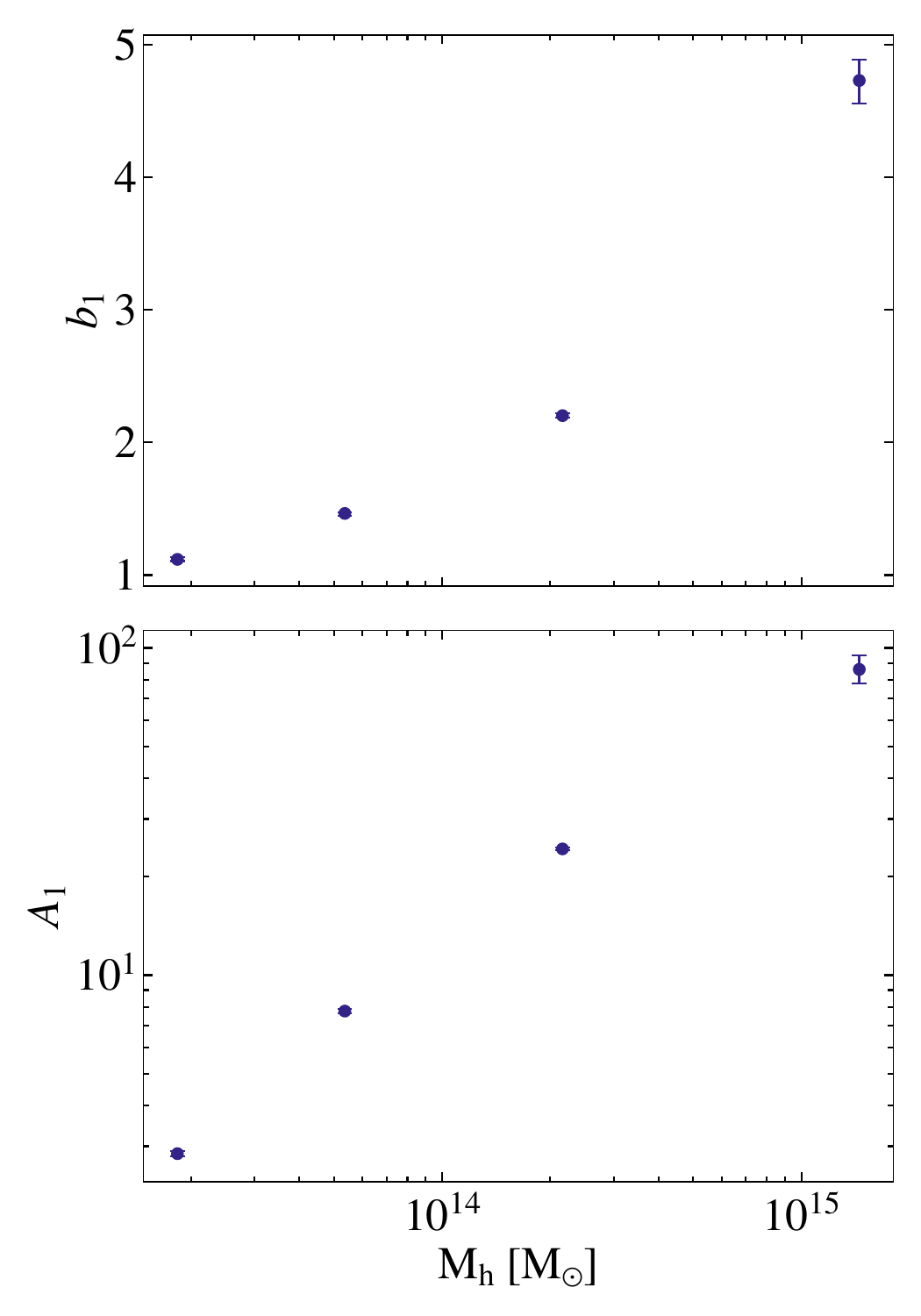}
    \caption{Variation of the non-linear bias parameters $b_1$ and the NLA alignment amplitude $A_1$ with halo mass $M_{\mathrm{h}}$ (in our case, we use $\mathrm{M}_{200\mathrm{mean}}$ as the mass of the halo) for our galaxy sample. $b_2$ is not shown for the sake of clarity. Error bars on the bias parameter and amplitude were calculated by taking the 68th percentile values of the IA posteriors.}
    \label{fig:nla_halo_mass}
\end{figure}

\begin{figure}
    \centering
    \includegraphics[scale=0.52]{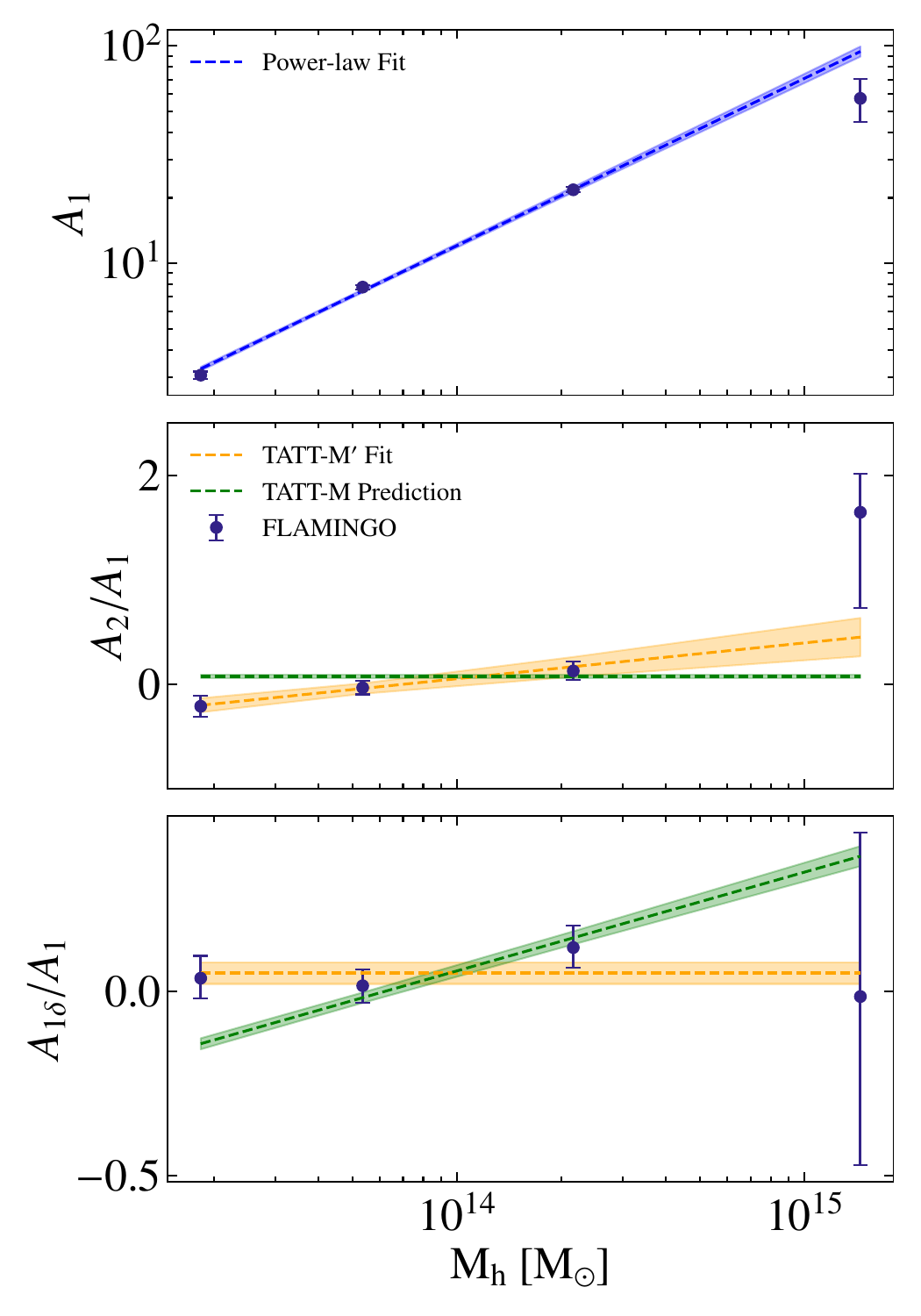}
    \caption{Variation of the TATT alignment amplitudes $A_1$, $A_2$ and $A_{1\delta}$ with halo mass $M_{\mathrm{h}}$ (in our case, we use $\mathrm{M}_{200\mathrm{mean}}$ as the mass of the halo) for our central galaxy sample. Error bars on the bias parameter and amplitude were calculated by taking the 68th percentile values of the IA posteriors. Overplotted in green is the prediction for $A_2$ and $A_{1\delta}$ from our TATT-M model fit with error bars. Informed by the fits to the halo sample (Fig. \ref{fig:TATT-M_ITs}), we assume the relations in Eqns. \ref{eqn:a2_a1} and \ref{eqn:a1d_a1} for the galaxy sample. In orange is the fit to the galaxy sample directly, which we call TATT-M$^\prime$.}
    \label{fig:tatt_halo_mass}
\end{figure}

\section{Dependence on mass}
\label{sect:mass-dependence}

Since alignment is a gravitational effect, it is natural that its strength depends on mass. \citet{Schneider2012} provided some of the first strong evidence for this using data from the Millenium and Millenium-2 simulations. They attributed this increase of alignment with halo mass to the fact that higher mass haloes formed earlier and hence are more biased. Several observational studies also explored the dependence on mass, or its more observationally accessible proxy, luminosity. The alignment amplitude has also been shown to increase with galaxy luminosity, the more observationally accessible proxy of mass (covered in Section \ref{sect:comparison}). \citet{Hao2011} find that the BCG alignment increases with stellar mass for clusters in the Sloan Digital Sky Survey (SDSS) DR7. This trend has even been shown to extend out to $\mathrm{M}_{200}$ masses of $\sim 10^{15} \ \mathrm{M}_{\odot}$ in the SDSS DR8 cluster sample \citep{vanUitert2017a}. \citet{Piras2018} improved on the simulation side by using the gravity-only Millenium-XXL simulations, and showed that the alignment amplitude scales with the halo mass as a power law with an index, $\beta_M$, between 1/3 and 1/2. 

\citet{Fortuna2021b} extended the luminosities probed using the KiDS-1000 sample of LRGs, which contained fainter galaxies (and hence lower mass) than previous analyses, and found that a broken power law was necessary to describe the dependence of $A_1$ on luminosity in the \textit{r}-band. Using the PAUS data, this was further extended to even fainter galaxies \citep{Navarro-Girones2025}. \citet{Fortuna2025} extended their previous analysis by linking luminosities with halo mass and found that a single power law of the form

\begin{equation}
\label{eqn:mass}
    A_1(M) = \alpha_{\mathrm{M}} \bigg( \frac{M}{M_0} \bigg)^{\beta_{\mathrm{M}}},
\end{equation}

\noindent was a good fit to their data. Here $\alpha_{\mathrm{M}}$ and $\beta_{\mathrm{M}}$ are the amplitude and slope of the power-law respectively, and $M_0 = 10^{13.5} \ \mathrm{M}_{\odot}$ is the pivot mass. This model was used as the fiducial IA model in the final KiDS-Legacy analysis \citep{Wright2025}, and was dubbed NLA-M because of the explicit mass dependence.

\subsection{Mass dependence of NLA-$A_1$}

We split the galaxy sample from \pkg{FLAMINGO} into halo mass bins and measured $w_{\mathrm{gg}}$ and $w_{\mathrm{g+}}$ in each bin, with a $\Pi_{\mathrm{max}}$ of 20 Mpc/$h$. We use the $\mathrm{M}_{200\mathrm{mean}}$ mass whenever we refer to halo mass, $M_{\mathrm{h}}$. The resulting best-fitting parameters with 1$\sigma$ error bars are shown in Fig. \ref{fig:nla_halo_mass}. We show the variation of the non-linear galaxy bias parameter $b_1$ and the alignment amplitude $A_1$ for an NLA fit as a function of halo mass. We see that $A_1$ increases as a power-law of mass, and that $b_1$ also grows with mass \citep[this has been studied extensively in the literature, for example in ][and we do not comment on it here]{Tinker2010}. The $b_2$ term is not shown here for clarity. When binning in mass, we have fewer samples for the modelling step than with the full sample, which causes the degeneracy between $b_1$ and $b_2$ to be exacerbated. We describe this problem in more detail in Appendix \ref{appendix:pimax_convergence}. We set a Gaussian prior on $b_2$ with mean 0.04 and standard deviation of 0.2 to solve this issue, informed by the posterior of the $b_2$ fit to the full galaxy sample but by scaling the standard deviation in accordance with the difference in sample sizes. We checked that even with a broad uniform prior on $b_2$, the constraints on NLA-$A_1$ remain unchanged except for a large scatter in the highest mass bin which had the lowest number of galaxies. We also did not observe any clear trend with mass for $b_2$.

\subsection{Mass dependence of TATT parameters}

In Fig. \ref{fig:tatt_halo_mass} we show for the first time the variation of the TATT parameters with halo mass ($\mathrm{M}_{200\mathrm{mean}}$). Similarly to Fig. \ref{fig:nla_halo_mass}, we jointly model the clustering and alignment (with TATT) signals for galaxies binned in halo mass. We show the variation of $A_1$, $A_2/A_1$ and $A_{1\delta}/A_1$ with halo mass. Since the focus of this work is on the alignment parameters, the variations of the two non-linear bias parameters $b_1$ and $b_2$ from the clustering signal are not shown. We also set a prior on $b_2$ to reduce the impact of the degeneracy between $b_1$ and $b_2$ for the last mass bin which contains few samples.

TATT improves upon NLA by including higher order terms that capture the torquing of the angular momentum and the density-weighting, thereby introducing two new terms $A_2$ and $A_{1\delta}$. Weak lensing analyses, however, rarely exploit the relation between these parameters as they set the same uniform prior ranges on the higher order TATT parameters $A_2$ and $A_{1\delta}$ (or equivalently $b_{\mathrm{TA}}$) as for the first order $A_1$ parameter \citep[see for example Table II of ][]{Abbott2025}. From Fig. \ref{fig:tatt_halo_mass} we see that these higher order terms are roughly an order of magnitude smaller than the corresponding $A_1$ in that bin. This motivates the use of more constraining priors on the higher order terms of TATT in future cosmic shear analyses, as one would naturally expect the higher order terms to be smaller than the leading order terms (if not, this would in fact motivate including even higher order terms in the analysis).

\subsection{TATT-M}

Similar to previous studies \citep{Fortuna2021b, Fortuna2025}, we found that a single power law fits the variation of $A_1$ with mass, shown in blue in the top panel of Fig. \ref{fig:tatt_halo_mass}. The best-fitting value of $\alpha_{\mathrm{M}}$ is $6.54^{+0.106}_{-0.118}$ and $\beta_{\mathrm{M}}$ is $0.77^{+0.015}_{-0.015}$. Given the size of the 2.8 Gpc$^3$ box, we had many more well resolved haloes than galaxies. We exploited this to get better statistics by splitting our halo sample (with bound dark matter mass more than $10^{12} \ \mathrm{M}_{\odot}$) into 8 mass bins in $\mathrm{M}_{200\mathrm{m}}$. For our halo sample, $A_2$ is roughly a constant multiple of $A_1$, and $A_{1\delta}/A_1$ increases as a power-law with mass. We found that $A_2$ and $A_{1\delta}$ can be represented as functions of halo mass as

\begin{equation}
\label{eqn:a2_a1}
    A_2 = k_1 A_1
\end{equation}

\noindent and 
\begin{equation}
\label{eqn:a1d_a1}
    A_{1\delta} = A_1(k_2 \mathrm{log_{10}}(M_{\mathrm{h}}/M_0)-k_3)
\end{equation}

\noindent where $k_1=0.073^{+0.015}_{-0.015}$, $k_2=0.268^{+0.016}_{-0.016}$ and $k_3=-0.038^{+0.014}_{-0.014}$. We set the same pivot mass $M_0$ as \citet{Fortuna2021b} for consistency and as this value was very close to the mean halo mass of our central galaxy sample. We arrive at the fits of the relations of $A_2$ and $A_{1\delta}$ with $A_1$ empirically from the simulation. In the case of $A_{1\delta}$, there is theoretical motivation for the functional form chosen in Eqn. \ref{eqn:a1d_a1}. Since the intrinsic shape field is measured only in biased locations, $C_{1\delta}$ must be related to the linear galaxy bias. \citet{Blazek2011} use this argument to motivate the relation $C_{1\delta} = b_1 C_1$. Using a shape-bias expansion method, \citet{Akitsu2023} showed that $A_{1\delta}/A_1 \propto (b_1 - k)$ where $k$ is some constant. Since the linear galaxy bias is a function of mass, $A_{1\delta}/A_1$ will also be mass-dependent, with a functional form like in Eqn. \ref{eqn:a1d_a1}. A full theoretical derivation of these relations is beyond the scope of this work, and we continue to use the empirically derived relations. 

These values were calculated for the simple iterative inertia tensor scheme. The exact values will change based on the inertia tensor scheme used, but the functional form of these equations will still be valid. We show the robustness of this model to the choice of inertia tensor scheme in Appendix \ref{appendix:inertia_tensors}.

Using the relations thus derived from the central halo sample, we predicted the $A_2$ and $A_{1\delta}$ values for the central galaxy sample from the best-fitting $A_1$ and the mean halo mass for each halo mass bin. This is shown in green in Fig. \ref{fig:tatt_halo_mass}. We show 1$\sigma$ errors on the prediction by sampling from Gaussians based on our best-fitting values, and taking the median, 16th and 68th percentiles from 5000 realisations. Moreover, the scaling of the amplitudes of $A_2$ and $A_{1\delta}$ between the halo sample (for which we fit the equations), and the galaxy sample (for which we make the prediction) is taken into account automatically by the change in $A_1$. This simple empirically derived fit works reasonably well for the galaxy sample, and the deviations from the predictions in the highest mass bin may be due to poor statistics. Since there are few objects in these bins, the fitting is more susceptible to parameter degeneracies. We also compared this TATT-M model from the halo sample with one based on fits to the galaxy sample, and find that the halo sample version works better.

\begin{figure}
    \centering
    \includegraphics[scale=0.3]{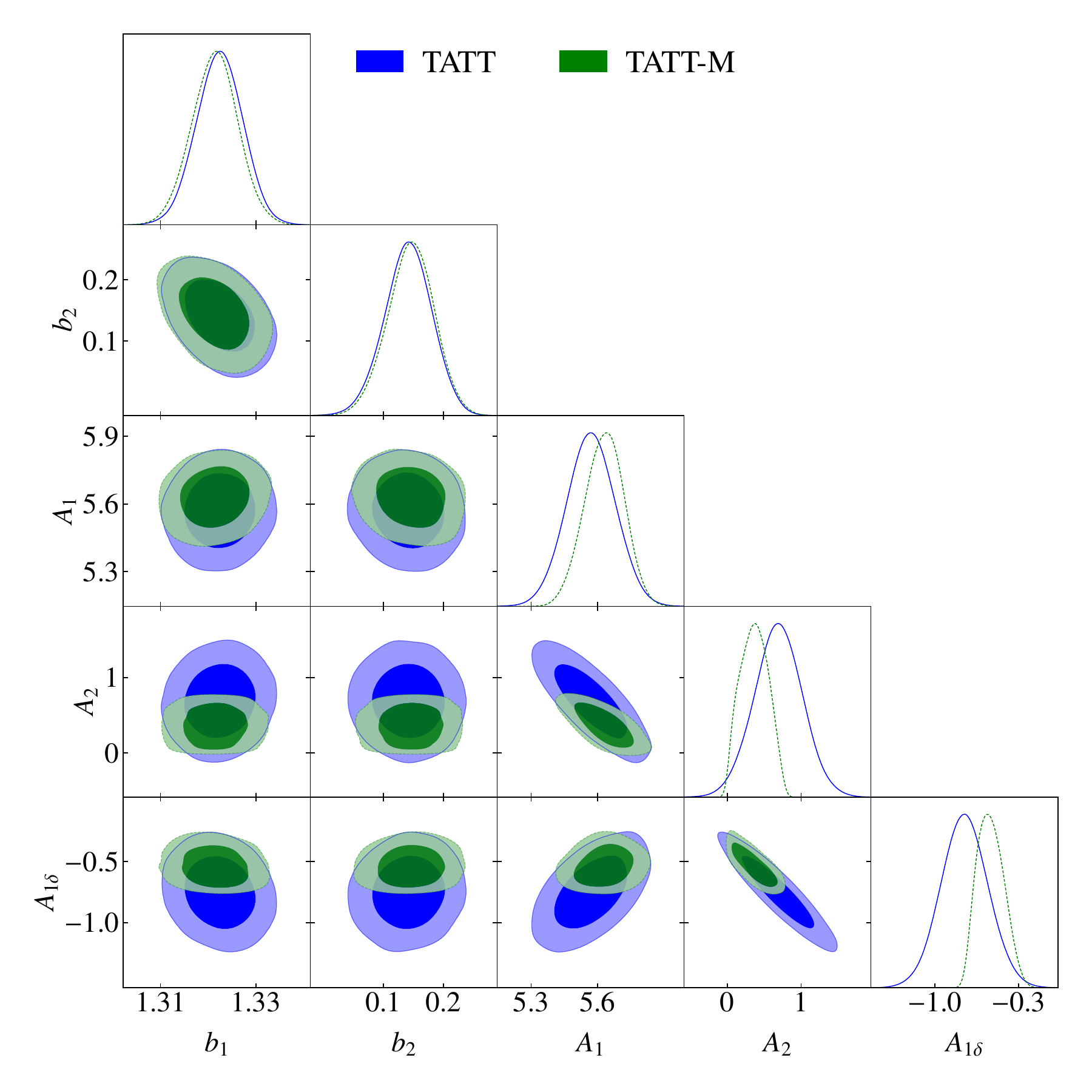}
    \caption{Corner plot of the posterior for the joint fit of galaxy clustering using non-linear bias and alignment signal using two models: TATT and TATT-M; over the scales 5-100 Mpc/$h$. TATT-M exploits empirically derived fitting functions between the TATT parameters and halo mass. The galaxy sample consists only of central galaxies, and the sampling was done by fitting $b_1$, $b_2$, $A_1$, $A_2$ and $A_{1\delta}$ for TATT, and for TATT-M by replacing $A_2$ and $A_{1\delta}$ with $k_1$, $k_2$ and $k_3$. These three extra parameters are sampled from the posterior of the fits to our halo sample, for which we have better statistics than for the galaxy sample. Due to the large amount of prior information, the overall constraining power is higher with our TATT-M model than with TATT. The posteriors on $A_2$ and $A_{1\delta}$ for TATT-M are derived using the relations in Eqns. \ref{eqn:a2_a1} and \ref{eqn:a1d_a1}. Note that these were not sampled and are shown here only for comparison.}
    \label{fig:tatt_vs_tattm_corner}
\end{figure}

\begin{figure}
    \centering
    \includegraphics[width=\columnwidth]{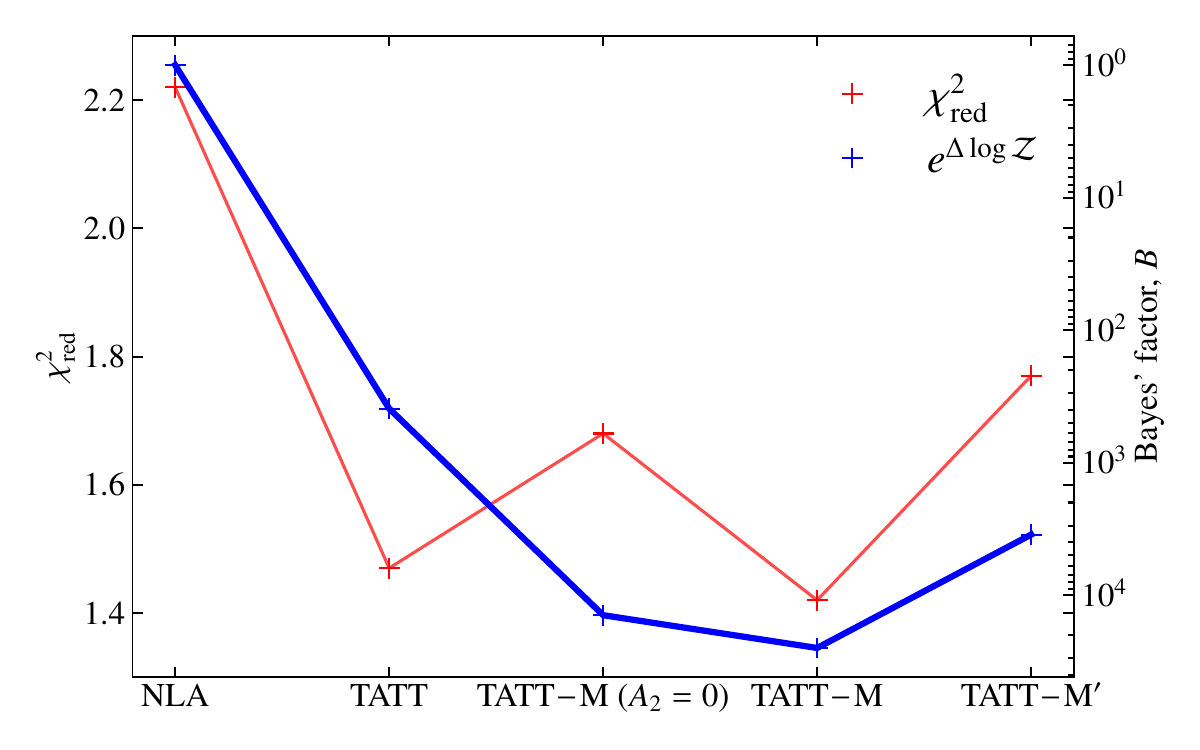}
    \caption{The values of the reduced chi-squared ($\chi_{\mathrm{red}}^2$) and the Bayes' factor relative to NLA for TATT, TATT-M ($A_2$ = 0), TATT-M and TATT-M$^\prime$. Our TATT-M model, based on an empirical fit to the halo data, is very strongly preferred over NLA on the central galaxy sample. TATT, TATT-M ($A_2$ = 0) and TATT-M$^\prime$ (based on empirical fits to the galaxy sample), are also preferred over NLA. There is a strong preference for the TATT-M model.}
    \label{fig:model_comparison}
\end{figure}

This paves the way for the use of our model, which we call TATT-M, in the same way as NLA-M was used in the KiDS-Legacy analysis \citep{Wright2025}. We effectively reduce TATT to a single parameter, $A_1$, that needs to be determined from the data. $A_2$ and $A_{1\delta}$ can be set from this derived $A_1$ value, and for the case of $A_{1\delta}$ with the addition of the halo mass. In practice, each tomographic bin in a cosmic shear analysis will have a luminosity associated with it, which can be used to derive a halo mass for that bin assuming a halo mass-luminosity relation. We can further constrain the amplitude by setting a power-law scaling for the alignment amplitude within each tomographic bin as a function of halo mass \citep{Fortuna2025}.

Note that since $k_2$ and $k_3$ are highly correlated in our formalism, in practice we employed a Principal Components Analysis (PCA) decomposition to the posteriors of $k_1$, $k_2$ and $k_3$ from the fit on the halo sample. We then sampled over the resulting decorrelated components to fit the TATT-M model to our central galaxy sample. Essentially, the two alignment parameters $A_2$ and $A_{1\delta}$ are replaced by the three TATT-M fit parameters $k_1$, $k_2$ and $k_3$. These parameters have very informative priors from the fit to the halo sample, so even though we increased the parameter space, the constraining power of the overall fit was increased. This is in-line with what was observed in \citet{Wright2025} for the NLA-M model.

We show a comparison of our TATT-M model against the base TATT in Fig. \ref{fig:tatt_vs_tattm_corner} fit to measurements of $w_{\mathrm{gg}}$ and $w_{\mathrm{g+}}$ on the central galaxy sample with $\Pi_{\mathrm{max}}=100$ Mpc/$h$. Using the posteriors of $k_1$, $k_2$ and $k_3$ and the mean halo mass of the sample ($\mathrm{log}_{10}M_{\mathrm{h}}/\mathrm{M}_{\odot}=13.647 $), we derived the posteriors on $A_2$ and $A_{1\delta}$. The posteriors of $k_1$, $k_2$ and $k_3$ are not shown separately as they are implicitly shown in the posteriors of $A_2$ and $A_{1\delta}$ of the TATT-M model. The posteriors of the alignment amplitude $A_1$ become smaller when using TATT-M due to the reduction in effective parameter space. The ratio of the error bars for the marginalised $A_1$ between TATT-M and TATT is 0.27. Thus, there is an increase in constraining power on $A_1$ between TATT and TATT-M by assuming the highly informative prior on the fitting relations from the simulation.

From Fig. \ref{fig:tatt_halo_mass}, we see that the variations of $A_2/A_1$ and $A_{1\delta}/A_1$ do not overlap perfectly with the prediction from the haloes. The variations of these parameters from the galaxies alone seem to suggest a constant for $A_{1\delta}/A_1$ and a linear fit on mass for $A_2/A_1$. We also tested an alternate TATT-M model where the functional form for $A_2/A_1$ is a linear function of log($M_{\mathrm{h}}$) and $A_{1\delta}/A_1$ is a constant, called TATT-M$^\prime$. Moreover, since the value of $A_2/A_1$ is quite close to 0, we also tested a model where the fit on the halo sample is used for $A_{1\delta}$ and $A_2$ is set to 0, called TATT-M ($A_2$ = 0). In Fig. \ref{fig:model_comparison} we compare NLA, TATT, TATT-M ($A_2$ = 0), TATT-M and TATT-M$^\prime$ using the same simulation data vector. We jointly fit these models with non-linear galaxy bias to $w_{\mathrm{gg}}$ and $w_{\mathrm{g+}}$ measured from the central galaxy sample from the 2.8 Gpc box. We calculated the $\chi_{\mathrm{red}}^2$ of the fits, as well as the Bayes' factor, which is the exponential of the difference between the Bayesian evidence of NLA and the rest of the models ($e^{\Delta \mathrm{log}(\mathcal{Z})}$). 

From the $\chi_{\mathrm{red}}^2$ values alone, we see that the TATT and TATT-M variants perform better than NLA. In the case of the mass dependant TATT models, the three fit parameters $k_1$, $k_2$ and $k_3$ have very informative priors as they were fit empirically to the simulation, which is accounted for in the Bayesian evidence calculation. These were calculated using {\tt nautilus} and are accurate to within 0.01 as our $N_{\mathrm{eff}} > 10 \ 000$. From this, it is clear that the data strongly prefers TATT and its variations over NLA. Moreover, since TATT has more freedom in its parameters than TATT-M and TATT-M$^\prime$, the latter models are preferred even more strongly. From this comparison, we also see that TATT-M describes the data better than TATT-M$^\prime$. Even though the prediction from the halo sample deviates at some masses from the galaxy sample (Fig. \ref{fig:tatt_halo_mass}), it performs quite well at the mean halo mass of the sample we tested these models on. This could explain why TATT-M works better than the TATT-M$^\prime$ model. Moreover, setting $A_2$ to 0 results in a worse $\chi_{\mathrm{red}}^2$ compared to TATT, but a better Bayes' factor, as there is one less parameter to fit. The inclusion of the mass-dependant $A_2$ in TATT-M improves both the $\chi_{\mathrm{red}}^2$ and the Bayes' factor. Overall, the TATT-M model (based on fits to the halo sample) best describes the data.

\section{Dependence on feedback}
\label{sect:feedback}

A limited number of studies have looked into the effect of feedback on alignments, given the cost of running simulations with different feedback modes. \citet{Velliscig2015a} used the feedback variations of the Cosmo-OWLS \citep{LeBrun2014} and EAGLE \citep{Schaye2015} simulation suites to show that the misalignment angle between the stellar component and the dark matter varies with feedback \citep[see also][]{Herle2025}. \citet{Tenneti2017} used the MassiveBlack-II \citep{Khandai2015} simulation to explore the effect of changing the sub-grid parameters controlling feedback on the alignment signal, and found their results were not sensitive to feedback. They did however, find a dependence of the misalignment angle on the feedback variation \citep{Velliscig2015a, Herle2025}. The main limitation of the work by \citet{Tenneti2017} is the small volume considered, as their simulation had a box size of only 25 Mpc/$h$. \citet{Soussana2020} explored the impact of feedback on intrinsic galaxy alignments in the Horizon simulations. They used two variations, one with AGN and the other with no AGN feedback, and found that this changes the fraction of galaxies that are pressure supported, thus changing the alignment statistics. Note that these represent two extreme scenarios for feedback. More recently, \citet{Bilsborrow2025} used the CAMELS simulation suite to detect a correlation between the alignment signal and the strength of supernova feedback. However, given that each box in their suite had a box size of only 25 $h^{-1}$ Mpc, their results were inconclusive for the effect of AGN feedback. 

The effect of feedback on intrinsic alignment parameters remains an open problem in the field. Broadly, feedback can affect the alignment of a galaxy in two main ways. First, a pressure-supported galaxy experiencing a tidal field in the early stages of its formation is sensitive to the distribution of the matter around it. AGN influence the matter field by injecting energy into the baryons, which then also influence the dark matter gravitationally. The redistribution of the matter in-turn changes the tidal field that the proto-galaxy experiences, thus changing its shape and alignment. The second more indirect way is that feedback only changes the fractions of the types of galaxies that are formed. If a certain feedback scenario resulted in a universe with fewer pressure-supported galaxies (ellipticals) and more discs, then this would suppress the alignment amplitude in that universe, as ellipticals are the major contributor to the alignment signal. The real effect of feedback in the universe could be a mixture of both these mechanisms. Alignment models used in the literature thus far make no allowance for variations due to feedback, which poses a serious problem in the analysis of upcoming surveys as they are sensitive to small-scale physics.

\begin{figure}
    \centering
    \includegraphics[scale=0.3]{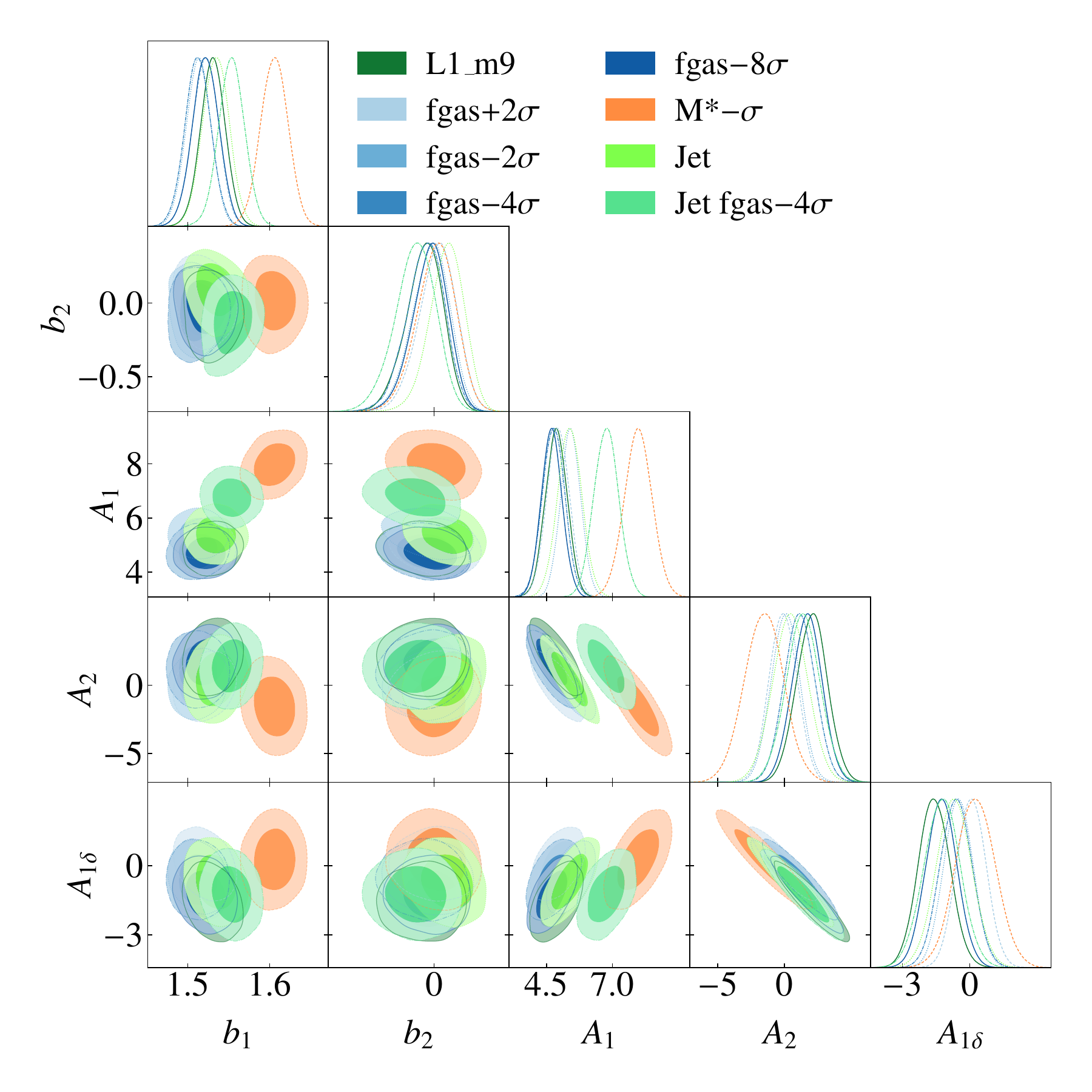}
    \caption{Corner plot of the posterior for the joint fits to $w_{\mathrm{gg}}$ with non-linear galaxy bias and $w_{\mathrm{g+}}$ with TATT for the different \pkg{FLAMINGO} feedback variations. The galaxy sample was selected from the 1 Gpc runs and consists only of objects with a dark matter mass of more than $10^{12} \ \mathrm{M}_{\odot}$ and more than 300 stellar particles. Feedback does not strongly change the derived constraints on $A_1$, except for the case of `M*-$\sigma$' shown in orange, and the strong jet variation `Jet fgas-4$\sigma$' shown in light green. These changes are consequences of the different stellar masses of the samples after the same stellar particle cut is applied.}
    \label{fig:naive_feedback_corner}
\end{figure}

The \pkg{FLAMINGO} simulations are well-suited to answer these questions for an LRG-like sample. Different feedback variations were implemented by re-tuning the sub-grid parameters of the simulation runs with the same initial conditions, resolution and cosmology \citep{Schaye2023, Kugel2023}. Model variations are defined in terms of their calibration data, which are shifted relative to their fiducial values, requiring different strength and modes of feedback. This results in a suite of feedback variations, as described in Table \ref{table:sim_details}.

Calibrating to the galaxy stellar mass function and cluster gas fractions from observations results in our fiducial feedback variation, called `L1\_m9'. Shifting the gas fraction by a certain number of standard deviations of the data results in `fgas+2$\sigma$', `fgas-4$\sigma$', etc. requiring simulations with varying strengths of AGN feedback, whereas shifting the stellar mass function (`M*-$\sigma$') mainly changes the strength of supernova feedback. We also use variations that implement a jet model for the AGN.

Numerous works have already used the feedback variations of \pkg{FLAMINGO} to test its effect on cosmological observables \citep[e.g.][]{McCarthy2023, McCarthy2025, Ondaro-Mallea2025}. Several of these that compared with observational studies found that the fiducial feedback variation is not strong enough to explain the data. For example, \citet{Siegel2025b} found that SDSS/DESI+ACT kSZ and eROSITA X-ray measurements prefer the strongest feedback scenario (`fgas-8$\sigma$') in \pkg{FLAMINGO}. However, pre-eROSITA X-ray cluster data are in tension with these stronger feedback scenarios \citep{Eckert2025}, and in fact prefer the fiducial feedback variation \citep{Braspenning2024}. Given that the discussion on the preference of the feedback variations from the \pkg{FLAMINGO} suite is still ongoing, we made use of all the feedback variations available. 

We created a galaxy sample for each feedback variation of the 1 Gpc box at redshift 0, with a dark matter mass of more than $10^{12} \ \mathrm{M}_{\odot}$ and more than 300 stellar particles. We fit non-linear galaxy bias and TATT jointly to $w_{\mathrm{gg}}$ and $w_{\mathrm{g+}}$ (with $\Pi_{\mathrm{max}} = 50$ Mpc/$h$), and the resulting posterior is shown in Fig. \ref{fig:naive_feedback_corner}. We account for the effect of feedback on the power spectra by using the power spectrum of each feedback variation for the modelling. We see that all models are consistent with each other within 1$\sigma$ for $A_1$, with the exception of `M*-$\sigma$' shown in orange. The strong jet variation (`Jet fgas-4$\sigma$') also results in an $A_1$ value that lies between the fiducial and the `M*-$\sigma$' case. The constraints on $A_2$ and $A_{1\delta}$ show much less variation with feedback. 

Naively, this would seem like an indication that feedback does in fact affect the alignment amplitude. However, the perceived change in $A_1$ in Fig. \ref{fig:naive_feedback_corner} represents the effect of the stellar particle cut imposed, as the `M*-$\sigma$' variation has fewer galaxies that have more than 300 star particles. The resulting galaxy sample has a larger mean halo mass, which increases the inferred $A_1$ value. When the effect of feedback on the stellar masses of a sample is not taken into account, we get results consistent with the findings of \citet{Bilsborrow2025}, who also find that stronger supernova feedback correlates with the alignment amplitude. In order to properly investigate the effect of changing baryon physics on the alignments, we need to remove the mass dependence.

\begin{figure}
    \centering
    \includegraphics[scale=0.4]{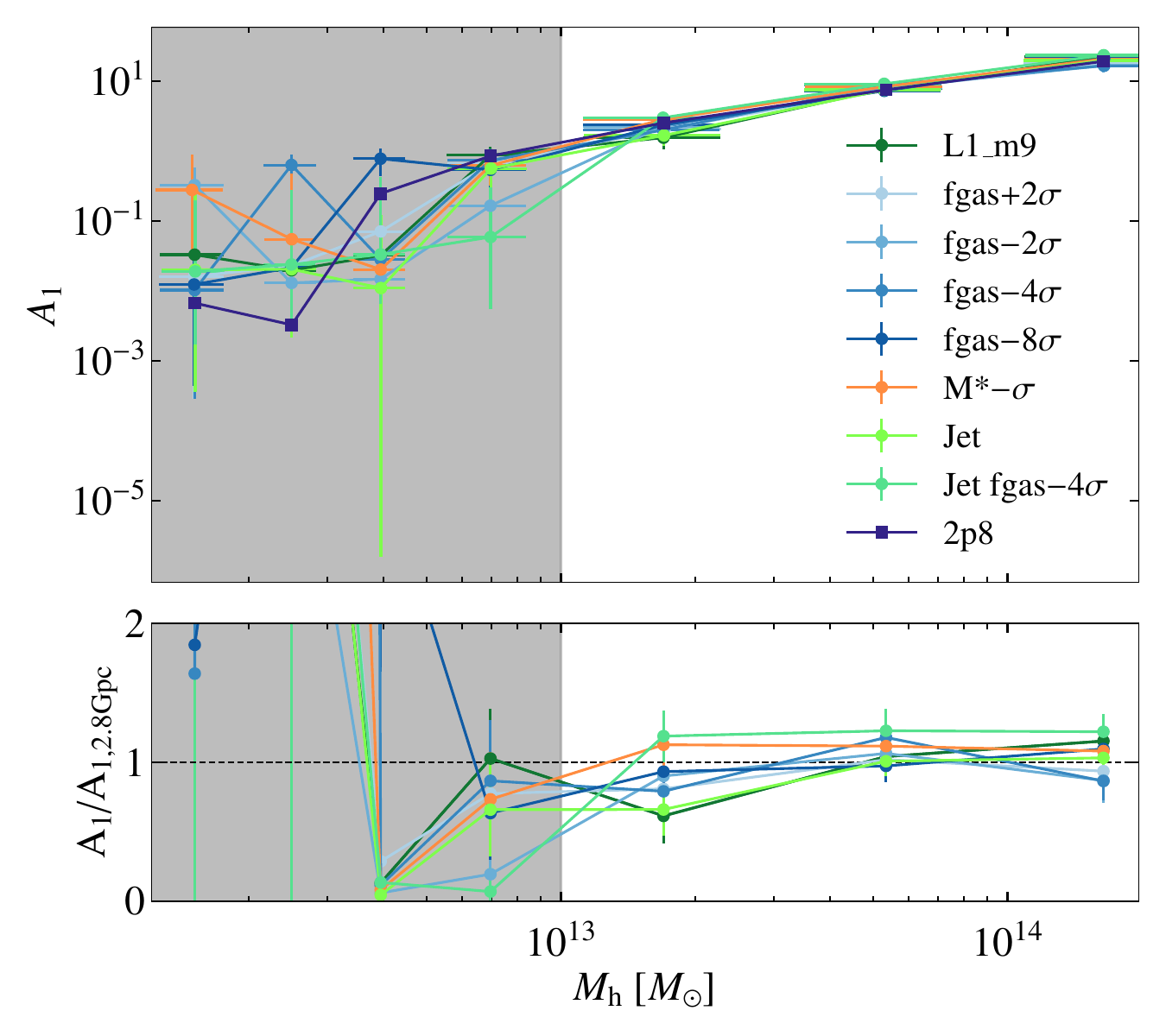}
    \caption{Top: Variation of TATT-$A_1$ with halo mass for each feedback model. The values of $A_1$ plotted are the best-fits with 1$\sigma$ errors in each halo mass bin. The error bars on the x-axis are the standard deviation of mass in each bin. Bottom: the ratio of each feedback variation with the values from the 2.8 Gpc, which is the fiducial feedback variation. The grey region represent data points below a halo mass of $10^{13} \ \mathrm{M}_{\odot}$, which are excluded from the fits for Fig. \ref{fig:alpha_beta_feedback}.}
    \label{fig:A1_Mh_feedback}
\end{figure}

To do so, we binned all the galaxies in the simulation in halo mass ($\mathrm{M}_{200\mathrm{m}}$). We then model each bin with TATT and plotted the $A_1$ values in Fig. \ref{fig:A1_Mh_feedback}. The best-fitting and 1$\sigma$ errors for $A_1$ in each halo mass bin are shown in the top panel, with the error in halo mass representing the bin width. The bottom panel shows the ratio of the values in each mode with the values for the 2.8 Gpc box, which was run with the fiducial feedback implementation.

\begin{figure}
    \centering
    \includegraphics[width=\columnwidth]{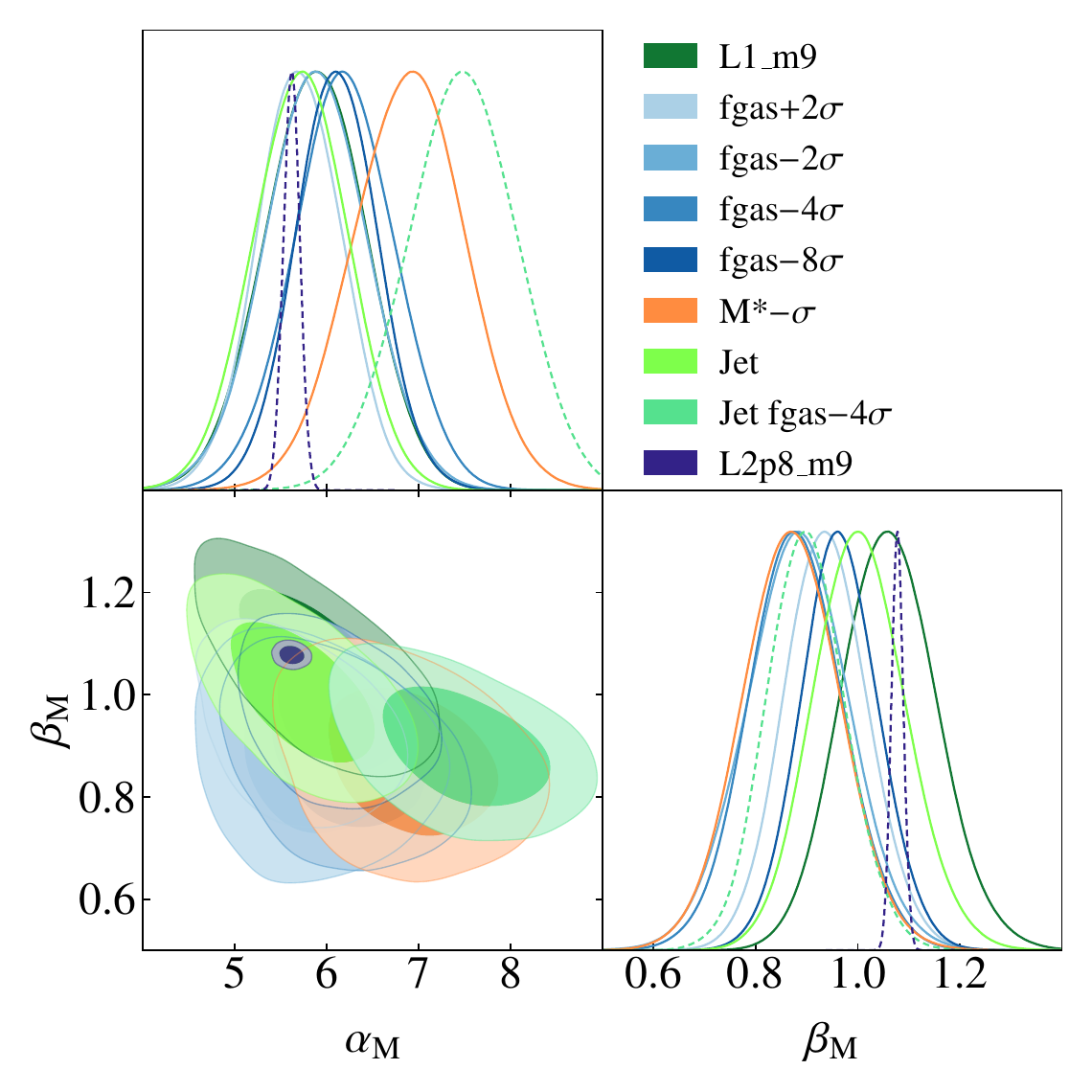}
    \caption{Corner plot of $\alpha_{\mathrm{M}}$ and $\beta_{\mathrm{M}}$ from Eqn. \ref{eqn:mass} for each of the \pkg{FLAMINGO} feedback models. These are posteriors from the fits on $A_1$ vs halo mass from Fig. \ref{fig:A1_Mh_feedback}. By fitting a power-law to the variation of $A_1$ with halo mass, the dependence on halo mass is removed, after which all the feedback modes become consistent with each other.}
    \label{fig:alpha_beta_feedback}
\end{figure}

Galaxies with fewer than 100 star particles were excluded from these measurements. Although these objects were excluded, the number removed varies significantly between feedback modes, which would have introduced biases similar to those seen in Fig. \ref{fig:naive_feedback_corner}. We evaluated the fraction of such poorly resolved galaxies in each mass bin and consequently removed the lowest four halo-mass bins with $M_{\mathrm{h}} < 10^{13} \ \mathrm{M}_{\odot}$ as they had $> 1$ per cent objects with fewer than 100 stellar particles. These are excluded from our analysis, and are shown in the grey region. Because we did not apply as strict a stellar-particle cut as in Fig. \ref{fig:naive_feedback_corner} ($N_*$ > 300), each halo-mass bin retains some contamination from shape noise, as a subset of galaxies still lack enough particles for a well-resolved shape. However, since all feedback modes should have the same shape bias and thus have the same error—and our primary goal is to compare the modes—we prioritised retaining the larger sample size for the analysis.

For the data-points with $M_{\mathrm{h}} > 10^{13} \ \mathrm{M}_{\odot}$, we fit a power-law as in Eqn. \ref{eqn:mass}. The posteriors for these fits are shown in Fig. \ref{fig:alpha_beta_feedback}. Once the mass dependence is removed, all the feedback modes are consistent with each other. Note also that the contours lie roughly around the posterior for the 2.8 Gpc box, with `L1\_m9' being completely consistent with `L2p8\_m9' as expected, as they have the same sub-grid implementation. 

The differences in $\alpha_{\mathrm{M}}$ and $\beta_{\mathrm{M}}$ between `M*-$\sigma$' or `Jet fgas-4$\sigma$' and the fiducial model are smaller compared to Fig. \ref{fig:naive_feedback_corner}. The small differences we see are statistically insignificant with the volume we have with the 1 Gpc$^3$ runs of \pkg{FLAMINGO}. Comparing the 1 Gpc$^3$ and 2.8 Gpc$^3$ volumes for the fiducial feedback scenario shows that they are consistent with each other, with much smaller error bars from the bigger box. If this trend were to hold for the other feedback variations, there could potentially be a very strong detection of the impact of feedback on the alignment signal. To truly test this, we would need larger boxes, comparable to the 2.8 Gpc$^3$ boxes, but for the different feedback scenarios. To probe lower halo masses we would need higher resolution simulations. From our analysis alone, there are no strong signs that feedback affects the alignment signal between galaxies with the same halo mass, but there are still indications that supernova and jet feedback can cause small variations in the alignment amplitudes. 

\section{Discussion and conclusions}
\label{sect:disc_conc}

The analysis of next generation survey weak lensing data will significantly reduce statistical uncertainties and necessitate a better understanding of one of the main astrophysical systematics in the data: intrinsic alignments. Given the large sky coverage and depth of such surveys, larger simulations with higher resolution than ever before are required to capture IA signals. Moreover, the interaction between baryons and dark matter needs to be incorporated into these simulations, which strongly motivates the need for hydro-dynamical simulations. In this work, we used the \pkg{FLAMINGO} simulations to explore two main open questions in the field, the mass dependence of IA and the impact of baryonic physics. 

We modelled the $w_{\mathrm{gg}}$ and $w_{\mathrm{g+}}$ data vectors jointly using non-linear bias for the clustering and both NLA and TATT for the alignment signal. We found that even for the very stringent test of these models given the precision of the 2.8 Gpc$^3$ run galaxy sample of $\approx$ 4.9 million galaxies, NLA and TATT were able to provide reasonable fits to the data, with a residual of less than 5 per cent over the scales considered (Fig. \ref{fig:2p8_fits}). Given that IA will constitute 10 per cent of the total shear signal, being able to model this to within a few percent is sufficient for upcoming surveys like \textit{Euclid} and LSST. We also found that TATT allows the minimum scale cut considered to be lowered compare to NLA, consistent with the picture that TATT captures more non-linear effects than NLA. 

This showed that even though there are IA models that are more accurate and capture more of the non-linear physics of alignments such as EFT \citep{Bakx2023, Chen2024}, TATT and NLA can be sufficient for the analysis of bigger datasets, like that available with \pkg{FLAMINGO}, for two-point statistics. A comparison of models such as EFT or HYMALAIA \citep{Maion2024} against TATT and NLA are beyond the scope of this work. With higher SNR measurements from three-point statistics, however, \citet{Gomes2026} and Vedder et al. (in preparation) found that TATT fails to represent that data well and that inclusion of the velocity-shear term, which relates the velocity and density fields, plays a significant role. 

Given the range of halo masses of our galaxy sample, we are able to use our constraints on IA models to provide guidance on the IA priors for 3x2pt analyses. By comparing our best-fitting alignment amplitude for NLA with various LRG samples in the literature (Fig. \ref{fig:aia_vs_L}), we showed that the alignments in \pkg{FLAMINGO} are comparable to observational studies. Given the resolution limits of the simulation, we did not probe the faint-end of the luminosity function, but did remarkably well at high luminosities. Like many works before, both in simulations and observations, we show that the alignment amplitude depends on mass, or its proxy, luminosity.

This mass dependence is the focus of Section \ref{sect:mass-dependence}, where we modelled both NLA and TATT in bins of halo mass (Figs. \ref{fig:nla_halo_mass} and \ref{fig:tatt_halo_mass}). We showed that the alignment amplitude $A_1$ (for both NLA and TATT) is well represented by a single power-law of halo mass \citep{Fortuna2025}. The mass dependence of the shape bias parameters have been studied previously \citep{Akitsu2023, Maion2025}, but in this work we showed the dependence of the TATT higher order alignment terms $A_2$ and $A_{1\delta}$ on halo mass for the first time. From this, it became apparent that $A_2$ and $A_{1\delta}$ are both an order of magnitude smaller than the $A_1$ parameter, which potentially allows the prior ranges for these higher order terms to be reduced in future weak lensing analyses. Moreover, $A_2$ may be expressed as a constant scaling of $A_1$, and $A_{1\delta}/A_1$ may be expressed as a linear relation in logarithmic halo mass bins.

From this insight, we introduced the mass-dependent TATT model, TATT-M which is the higher-order equivalent of the NLA-M model that has been used successfully in the KiDS-Legacy analysis \citep{Fortuna2025, Wright2025}. By empirically fitting these relations to the halo sample for better statistics compared to the galaxy sample, we showed that they can be used to predict the higher order TATT terms for the galaxy sample remarkably well (Fig. \ref{fig:tatt_halo_mass}). Also, our model is robust to the choice of inertia tensor scheme (Fig. \ref{fig:TATT-M_ITs}). This paves the way for an implementation of TATT-M in a cosmic shear survey similarly to the implementation of NLA-M. Given a luminosity in each tomographic bin, a mean halo mass can be derived assuming a certain luminosity-halo mass relation and associated scatter. This is then folded into the IA modelling by including the index of the power law (Eqn. \ref{eqn:mass}) that relates the $A_1$ to the mean halo mass of the sample. Replacing $A_2$ and $A_{1\delta}$ by the fit parameters $k_1$, $k_2$ and $k_3$ allows the overall constraining power to increase given the extremely informative prior on these parameters from the simulation. The prior derived from simulations can be used to inform that used in an observational analyses (or used directly), or can be directly fit to an LRG sample from the survey data, similar to the analysis in \citet{Fortuna2025}. We applied our TATT-M model to our galaxy sample and compared it to TATT (Fig. \ref{fig:tatt_vs_tattm_corner}), and found that the constraining power on $A_1$ increases. We also show in Fig. \ref{fig:model_comparison} that our TATT-M model is strongly preferred over NLA for our central galaxy sample.

In Section \ref{sect:feedback} we explored the effect of baryonic physics on the alignment signal under TATT. In Fig. \ref{fig:naive_feedback_corner} we showed a simple test of the effect of feedback, by making the same stellar particle number cut of 300 (in order to have well resolved stellar shapes) on all the feedback variation runs. We found that the `M*-$\sigma$' and `Jet fgas-4$\sigma$' \pkg{FLAMINGO} variations differed the most from the fiducial feedback run. However, this difference could be entirely due to the different halo mass distributions of the samples after the same particle cut has been applied to the different feedback variations. We re-did this analysis after taking this halo mass dependence into account by binning in halo mass. We found that the amplitude and slope of the single power law is consistent between all the feedback modes within the error bars (Fig. \ref{fig:alpha_beta_feedback}). 

Within the resolution limits of \pkg{FLAMINGO}, there are no signs that feedback impacts the alignment signal beyond its effect on the galaxy stellar mass. However, this may change at smaller scales than we can probe with this simulation, and for lower halo mass objects (although higher mass haloes experience stronger feedback and may be more affected by it). 

We would require higher resolution simulations with more sophisticated galaxy formation physics to extend this work. Higher resolution would allow us to extend the mass dependence analysis down to lower masses. We would also be able to probe the effects of feedback on low mass haloes and at smaller scales than accessible with \pkg{FLAMINGO}. Moreover, the redshift evolution of the alignment models used in this work are still an open quesion. An ideal simulation to explores these topics further would be the \pkg{COLIBRE} suite of simulations \citep{Schaye2025, Chaikin2025a}, which has more sophisticated galaxy formation physics, which will be the focus of future work.

\begin{acknowledgements}
AH thanks Dennis Neumann, Casper Vedder, Guadalupe Cañas Herrera and Rob McGibbon for valuable discussions. AH acknowledges support by NWO through the Dark Universe Science Collaboration (OCENW.XL21.XL21.025). 

NEC acknowledges support from the project ``A rising tide: Galaxy intrinsic alignments as a new probe of cosmology and galaxy evolution'' (with project number VI.Vidi.203.011) of the Talent programme Vidi which is (partly) financed by the Dutch Research Council (NWO). 

HH and DNG acknowledge funding from the European Research Council (ERC) under the European Union's Horizon 2020 research and innovation program (Grant agreement No. 101053992).

This work used the DiRAC@Durham facility managed by the Institute for Computational Cosmology on behalf of the STFC DiRAC HPC Facility (www.dirac.ac.uk). The equipment was funded by BEIS capital funding via STFC capital grants ST/K00042X/1, ST/P002293/1, ST/R002371/1 and ST/S002502/1, Durham University and STFC operations grant ST/R000832/1. DiRAC is part of the National e-Infrastructure. 

This work employed the packages \pkg{NUMPY} \citep{Harris2020}, \pkg{MATPLOTLIB} \citep{Hunter2007}, \pkg{SCIPY} \citep{Scipy2020}, \pkg{SWIFTsimIO} \citep{Borrow2020} and CCL \citep{Chisari2019}.

This work benefited from the resources and support provided by the echo-IA collaboration (https://echo-ia.org).

\end{acknowledgements}

%
%
\bibliographystyle{aa} 
\bibliography{main} 

\begin{appendix}
\section{Effect of changing inertia tensor}
\label{appendix:inertia_tensors}

\begin{figure}
    \centering
    \includegraphics[scale=0.5]{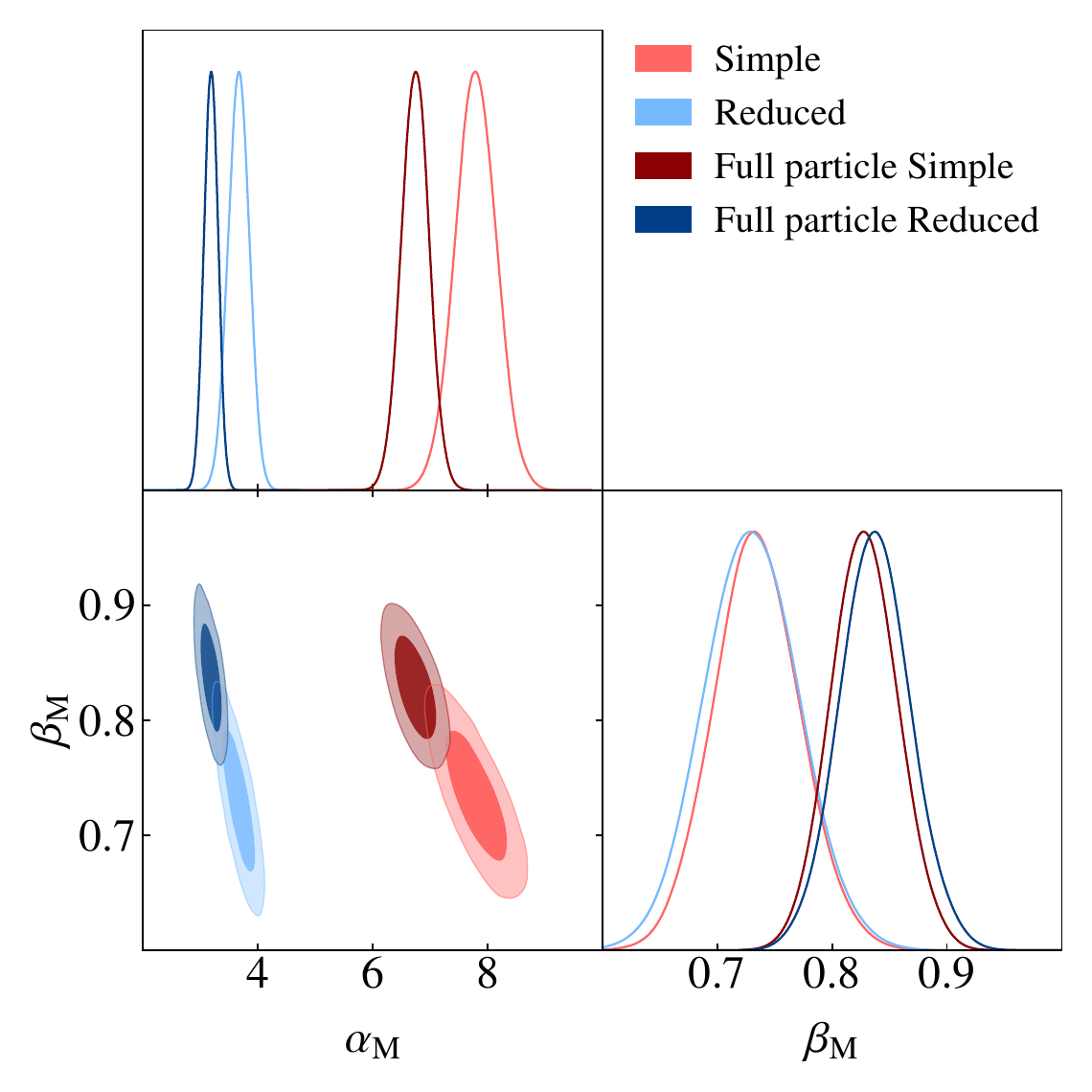}
    \caption{The posteriors for the best-fit power law for the variation between the alignment amplitude and halo mass (Eqn. \ref{eqn:mass}), for galaxies in the 1 Gpc fiducial run. Posteriors for the simple and reduced (iterative) stellar inertia tensors are shown, considering only the particles within the half-mass radius, and recomputed  considering all bound particles. The recomputed inertia tensors have a lower amplitude and steeper slope compared to the tensor calculated using only the particles within the half-mass radius.}
    \label{fig:alpha_beta_1Gpc}
\end{figure}

\begin{figure}
    \centering
    \includegraphics[scale=0.5]{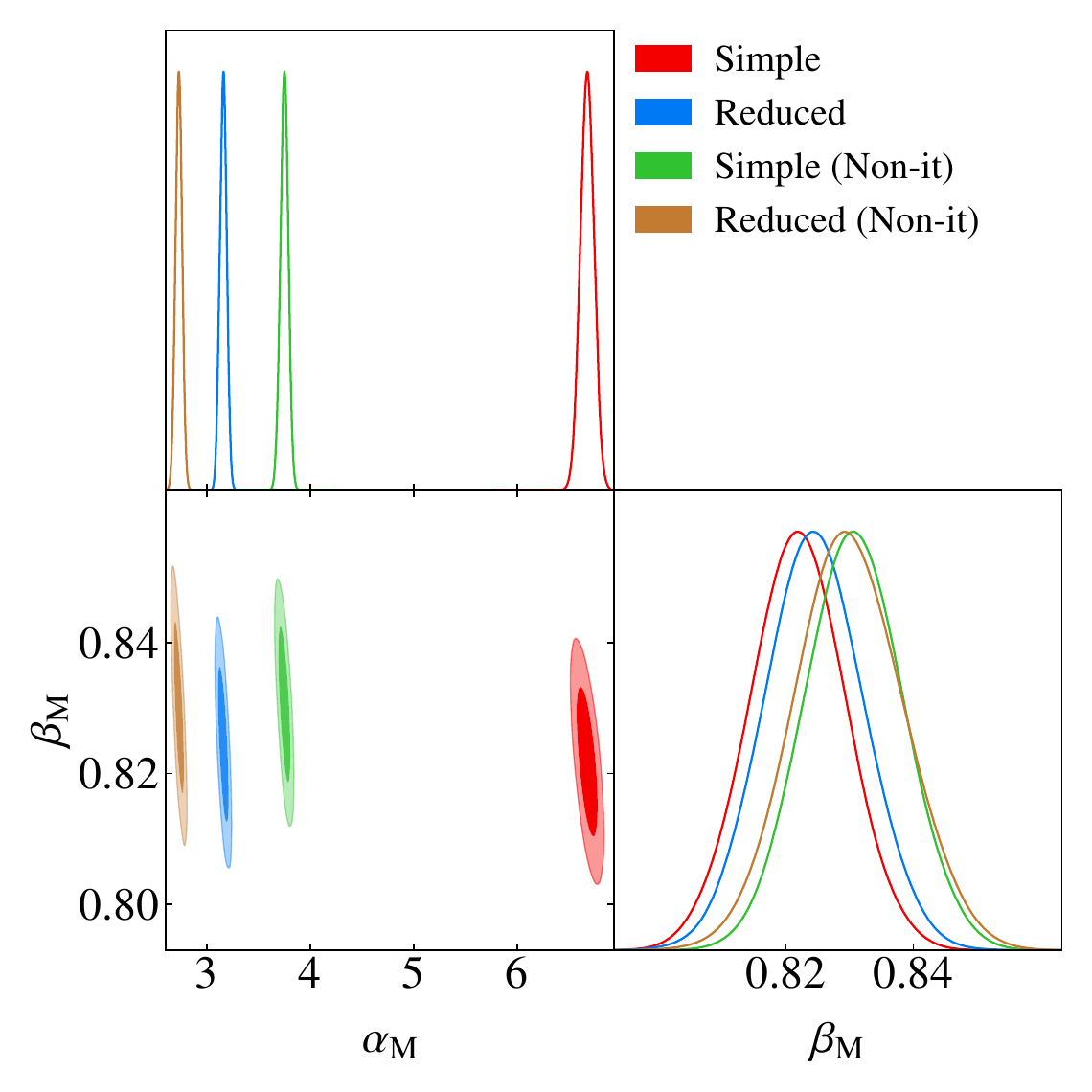}
    \caption{The posteriors for the best-fitting power law for the variation between the alignment amplitude and halo mass (Eqn. \ref{eqn:mass}), for galaxies in the 2.8 Gpc fiducial run. Posteriors for the simple iterative, reduced iterative, simple non-iterative and reduced non-iterative total inertia tensors are shown, considering only the particles within the half-mass radius. The effect of the choice of inertia tensor scheme on the slope is less than 1$\sigma$, and constitutes only an amplitude scaling.}
    \label{fig:alpha_beta_2p8Gpc}
\end{figure}

There are two main types of inertia tensors used in the literature. When particle positions are weighted by their inverse radius (down-weighting particles at larger distances, a method motivated by observational considerations) the resulting tensor is termed the reduced inertia tensor. In contrast, equal weighting of particle positions yields the simple inertia tensor. For both weighting schemes, the inertia tensor can be computed using either an iterative or non-iterative approach. In the non-iterative method, particles within a predefined spherical region are included in the calculation. The iterative method, however, replaces this bounding sphere with an ellipsoid whose shape is updated at each step based on the previously derived inertia tensor, continuing until convergence is achieved.  
\begin{table}
\caption{Best-fitting parameters and 1$\sigma$ errors for the relations between TATT-$A_1$ and mass (Eqn. \ref{eqn:mass}), and between $A_2$, $A_{1\delta}$ with $A_1$ and mass.}
\label{table:inertia_tensor_params}
\centering
\begin{tabular}{lcccc}
\hline
\noalign{\smallskip}
 & SIT & RIT & SIT  & RIT \\
 & & & Non-it. & Non-it. \\
\hline
\noalign{\smallskip}

$\alpha_{\mathrm{M}}$ &
$12.861^{+0.097}_{-0.098}$ &
$5.696^{+0.047}_{-0.045}$ &
$6.108^{+0.047}_{-0.046}$ &
$4.712^{+0.037}_{-0.039}$ \\[4pt]

$\beta_{\mathrm{M}}$ &
$0.461^{+0.004}_{-0.004}$ &
$0.550^{+0.004}_{-0.004}$ &
$0.424^{+0.004}_{-0.004}$ &
$0.536^{+0.005}_{-0.004}$ \\[4pt]

$k_1$ &
$0.073^{+0.015}_{-0.015}$ &
$0.061^{+0.018}_{-0.018}$ &
$0.065^{+0.014}_{-0.014}$ &
$0.069^{+0.017}_{-0.018}$ \\[4pt]

$k_2$ &
$0.268^{+0.016}_{-0.016}$ &
$0.270^{+0.020}_{-0.020}$ &
$0.264^{+0.016}_{-0.016}$ &
$0.240^{+0.019}_{-0.020}$ \\[4pt]

$k_3$ &
$-0.038^{+0.014}_{-0.014}$ &
$-0.038^{+0.017}_{-0.015}$ &
$-0.042^{+0.015}_{-0.013}$ &
$-0.056^{+0.016}_{-0.015}$ \\

\noalign{\smallskip}
\hline
\end{tabular}

\tablefoot{Values shown were calculated for the halo sample from the 2.8 Gpc box, for different inertia tensor schemes: simple iterative (`SIT'), reduced iterative (`RIT'), simple non-iterative (`SIT Non-it.'), and reduced non-iterative (`RIT Non-it.').}
\end{table}

Hence, there are four variations of the inertia tensors: simple iterative, simple non-iterative, reduced iterative and reduced non-iterative. All four definitions were implemented in SOAP \citep{McGibbon2025}. Compared to the iterative method, the non-iterative approach introduces a bias that artificially sphericises haloes, whereas the iterative procedure more accurately captures their true shapes \citep[see][for discussions]{Zemp2011, Bett2012, Velliscig2015a, Valenzuela2024}. The reduced tensors have lower weight at larger radii where the galaxies tend to be more strongly aligned, resulting in shapes that are less elliptical than the simple tensors. Since more elliptical objects will have a higher alignment amplitude, this means that the strength of the alignment amplitude, in order from highest to lowest, for the tensor schemes will be: simple iterative, simple non-iterative, reduced iterative and reduced non-iterative. 

Another detail to consider is which particles are included in the calculation of the inertia tensors. Typically in simulations, all bound particles are used for the calculation of the inertia tensors. However, in the case of the 2.8 Gpc box of \pkg{FLAMINGO}, including all the particles, especially for the most massive objects in the simulations, proved to be computationally intractable. Instead, only particles within the half-mass radius of each object were considered. In this appendix, we will discuss the effect of these choices on our results.

To explore the effect of excluding particles outside the half-mass radius, we use the `L1\_m9' run, which has a box size of 1 Gpc and was run using the fiducial feedback implementation, the same as used for the 2.8 Gpc run. For this smaller volume, the stellar inertia tensors were recomputed using all the bound particles in each object. We compare the simple and reduced inertia schemes calculated iteratively for two cases: one where only the particles within the half-mass radius were considered (our fiducial choice) and one where all bound particles were considered (`full particle'). We modelled the variation of the NLA-$A_1$ amplitude in halo mass bins (similar to Fig. \ref{fig:A1_Mh_feedback}) for these inertia tensor schemes. We fit the relation in Eqn. \ref{eqn:mass} to these data points, excluding the bins with $M_{\mathrm{h}} < 10^{13} \ \mathrm{M}_{\odot}$, and the resulting posteriors on $\alpha_{\mathrm{M}}$ and $\beta_{\mathrm{M}}$ are shown in Fig. \ref{fig:alpha_beta_1Gpc}. 

Including particles beyond the half-mass radius increases the power-law index by approximately 0.1 while reducing the amplitude. Because the amplitude is expected to rise at larger radii, the decrease we observe may reflect contamination from particles near the object’s boundary that were incorrectly assigned to it. We do not investigate this further, as the full inertia tensors are unavailable for the 2.8 Gpc box for historical reasons, and we prioritized the larger statistical power provided by its volume.

We repeated this exercise for the simple iterative, simple non-iterative, reduced iterative and reduced non-iterative tensors (only including particles within the half-mass radius) for the galaxies in the 2.8 Gpc box. We modelled the variation of $A_1$ (using NLA) for halo mass bins for the different inertia tensor types, and the corresponding power-law fit posteriors are shown in Fig. \ref{fig:alpha_beta_2p8Gpc}. As expected, the simple iterative scheme has the highest $\alpha_{\mathrm{M}}$ value, followed in order of decreasing $\alpha_{\mathrm{M}}$ by the simple non-iterative, reduced iterative and reduced non-iterative schemes. The power-law indices for each variation are also consistent with each other. This shows that the choice of inertia tensor simply changes the amplitude of the alignment signal, without affecting the mass and thus scale dependence. Thus, we preferred the simple iterative scheme as our fiducial scheme as it provides the highest amplitude, resulting in the highest signal-to-noise ratio .

\begin{figure}
    \centering
    \includegraphics[scale=0.5]{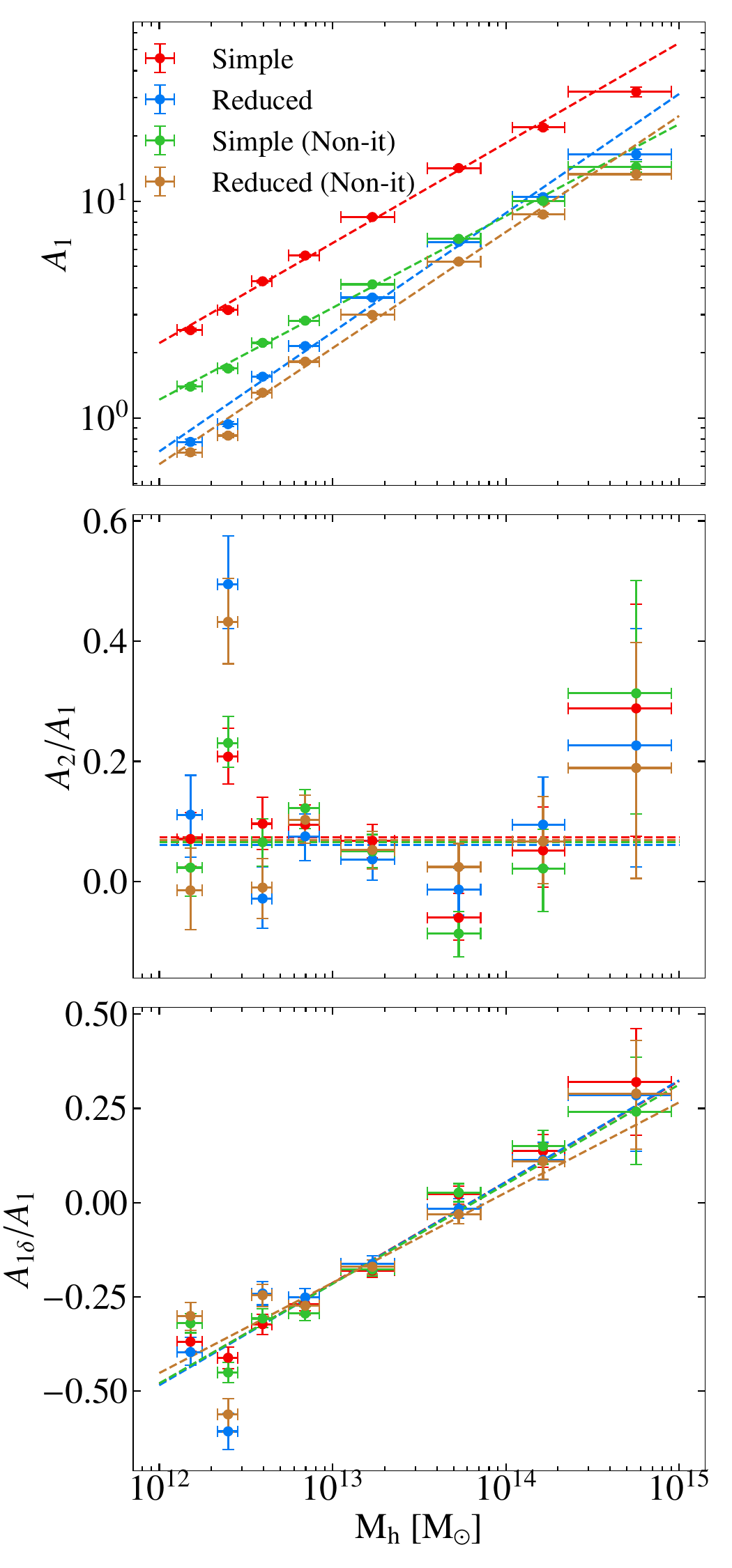}
    \caption{Variation of the TATT parameters for the halo sample of the 2.8 Gpc box with halo mass $M_{\mathrm{h}}$ (in our case, we use $\mathrm{M}_{200\mathrm{mean}}$ as the mass of the halo). The different lines correspond to the four types of inertia tensors. $A_1$ is a power-law of $M_{\mathrm{h}}$ irrespective of inertia tensor choice. Similarly, the functional forms for the variation of $A_2/A_1$ and $A_{1\delta}$ with mass for the TATT-M model hold for all choices of inertia tensors considered. We omit the error bars on the fits for clarity.}
    \label{fig:TATT-M_ITs}
\end{figure}

In Section \ref{sect:mass-dependence}, we introduced the TATT-M model for alignments. To derive the values of the fits for this model, we employed the simple iterative scheme. Fig. \ref{fig:TATT-M_ITs} shows the TATT parameters in halo mass bins for the four inertia tensor schemes. Here, we considered the inertia tensors calculated for the total matter in each object, with $M_{\mathrm{h}} > 10^{12} \ \mathrm{M}_{\odot}$. We can see that the functional forms we adopted for the relations between the TATT parameters hold for different choices of the inertia tensor scheme. As we saw in Fig. \ref{fig:alpha_beta_2p8Gpc}, only the amplitude $\alpha_{\mathrm{M}}$ changes significantly with this choice. Since any analysis adopting the TATT-M model will leave $A_1$ as a free-parameter, this change in amplitude will be captured based on the specific choice of the inertia tensor scheme. For reference, we include the best-fitting values of the power-law fits $\alpha_{\mathrm{M}}$ and $\beta_{\mathrm{M}}$, as well as the TATT parameters $k_1$, $k_2$ and $k_3$ in Table \ref{table:inertia_tensor_params}.

\section{Effect of changing $\Pi_{\mathrm{max}}$}
\label{appendix:pimax_convergence}

\begin{figure}
    \centering
    \includegraphics[scale=0.3]{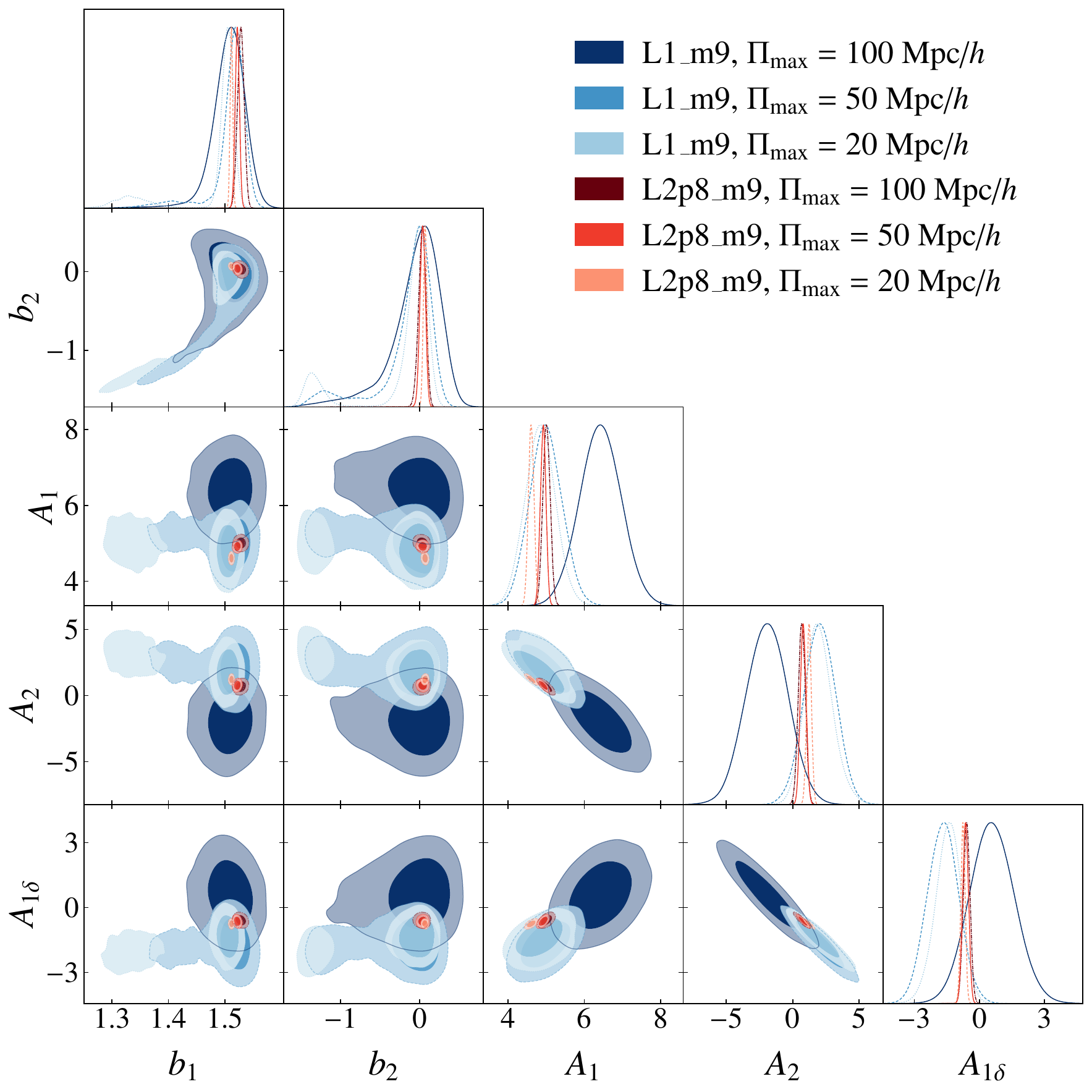}
    \caption{Posteriors of the joint fit of the clustering with non-linear bias and the alignment with TATT for different values of $\Pi_{\mathrm{max}}$, for the 1 Gpc box (in blue) and the 2.8 Gpc box (in red). The darker the shade of blue/red, the higher the value of $\Pi_{\mathrm{max}}$. Decreasing $\Pi_{\mathrm{max}}$ increases the SNR, and if there are sufficient statistics, different $\Pi_{\mathrm{max}}$ values give consistent answers. The degeneracy between $b_1$ and $b_2$ is exacerbated by smaller statistical power, and may even lead to biased results if too low a value of $\Pi_{\mathrm{max}}$ is chosen.}
    \label{fig:pimax_convergence}
\end{figure}

$\Pi_{\mathrm{max}}$ is the maximum line-of-sight separation integrated over in Eqn. \ref{eqn:wab}. Lower values of $\Pi_{\mathrm{max}}$ produce higher SNR measurements \citep[see][for an application of this]{Lamman2025}. Since we do not invoke the Limber approximation in Eqns. \ref{eqn:wgg} and \ref{eqn:wgp}, our inferred central value from the posteriors of the non-linear bias and alignment models should be unaffected by this choice. Typically, observations use fairly high values of $\Pi_{\mathrm{max}}$ in order to avoid contamination by redshift space distortions (RSD) \citep{Lamman2024b}. In this work, we have ignored RSD effects. 

In Fig. \ref{fig:pimax_convergence}, we show the effect of changing $\Pi_{\mathrm{max}}$ on the posteriors for the 2.8 Gpc$^3$ box. We also show posteriors from the 1 Gpc$^3$ box to test how lower statistics affects $\Pi_{\mathrm{max}}$ choice. As expected, the higher the value of $\Pi_{\mathrm{max}}$, the wider the contours. The contours from the fits on the 2.8 Gpc box are consistent with each other irrespective of the value of $\Pi_{\mathrm{max}}$. For the 1 Gpc box, the $\Pi_{\mathrm{max}}$ of the 100 Mpc/$h$ contour is consistent with the posteriors from the 2.8 Gpc box. At $\Pi_{\mathrm{max}}$ values of 50 and 20 Mpc/$h$ however, we begin to get biased values for $b_2$, which in turn affects $b_1$.

For the case of the 2.8 Gpc$^3$ volume, for which there are sufficient statistics, the posterior of $b_2$ is well behaved and centred on 0. When the statistical power is reduced, however, $b_2$ tends to go to negative values. Even though the focus of this work is the alignment signal, not the clustering signal, the degeneracy between $b_1$ and $b_2$ biases the inferred value of $b_1$, which in turn affects the value of the alignment amplitude. This was problematic for many fits in this work when statistics were lowered, for example when mass bins were created. In order to reduce this effect in fits where we have reduced statistics, we placed Gaussian priors on $b_2$ with mean 0 and standard deviation in proportion to the ratios of the number of objects considered. 

\end{appendix}

\end{document}